\documentclass[aps,prb,english,twocolumn,showpacs,preprintnumbers,amsmath,amssymb,floatfix,superscriptaddress,longbibliography]{revtex4-1}

\usepackage{chemformula} 
\usepackage[T1]{fontenc}
\usepackage[latin9]{inputenc}  
\usepackage{graphicx}
\usepackage{amsmath}
\usepackage{amssymb}
\usepackage{babel}
\makeatletter
\usepackage{siunitx}
\usepackage[version=3]{mhchem}
\usepackage{xcolor}
\usepackage{stmaryrd}
\usepackage{textcomp}
\usepackage{calrsfs}
\usepackage{yfonts}
\usepackage{bm}
\usepackage{multirow}
\global\arraycolsep=2pt 
\usepackage{hyperref}
\usepackage{adjustbox}
\usepackage[graphicx]{realboxes}
\hypersetup{
     colorlinks = true,
     citecolor  = blue, 
     urlcolor   = black, 
     linkcolor  = blue  
}

\begin{document}
\title{Exploring the wettability of liquid iron on refractory oxides with sessile drop technique and density-functional derived Hamaker constants}
\author{Sudhanshu Kuthe}
\email{kuthe@kth.se}
\affiliation{Department of Materials Science and Engineering, KTH Royal Institute of Technology, SE-100 44 Stockholm, Sweden}
\author{Mathias Bostr{\"o}m}
\affiliation{Centre of Excellence ENSEMBLE3 Sp. z o. o., Wolczynska Str. 133, 01-919, Warsaw, Poland}
\affiliation{Chemical and Biological Systems Simulation Lab, Centre of New Technologies, University of Warsaw, Banacha 2C, 02-097 Warsaw, Poland}
\author{Wen Chen}
\affiliation{UniversalLab GmbH, Park Innovaare: deliveryLAB, 5234 Villigen, Switzerland}
\affiliation{Yangtze Delta Region Institute of Tsinghua University, Zhejiang, 314006, China}
\author{Bj{\"o}rn Glaser}
\affiliation{Department of Materials Science and Engineering, KTH Royal Institute of Technology, SE-100 44 Stockholm, Sweden}
\author{Clas Persson}
\email{claspe@kth.se}
\affiliation{Department of Materials Science and Engineering, KTH Royal Institute of Technology, SE-100 44 Stockholm, Sweden}

\begin{abstract}
{The macroscopic interactions of liquid iron and solid oxides, such as alumina, calcia, magnesia, silica, and zirconia manifest the behavior and efficiency of high-temperature metallurgical processes.
The oxides serve dual roles, both as components of refractory materials in submerged entry nozzles and also as significant constituents of non-metallic inclusions in the melt. It is therefore crucial to understand the physicochemical interplay between the liquid and the oxides in order to address the nozzle clogging challenges, and thereby optimize cast iron and steel production. 
This paper presents a methodology for describing these interactions by combining the materials' dielectric responses, computed within the density functional theory, with the Casimir-Lifshitz dispersion forces to generate the Hamaker constants. 
The approach provides a comprehensive understanding of the wettability of iron against these refractory oxides, revealing the complex relation between molecular and macroscopic properties. Our theoretically determined crystalline structures are confirmed by room-temperature X-ray diffraction, and the contact angles of liquid iron on the oxides are validated with a sessile drop system at the temperature 1823\,K. For comparison, we also present the wettability of the oxides by a liquid tin-bismuth alloy.
The findings are essential in advancing the fundamental understanding of interfacial interactions in metallurgical science, and are also pivotal in driving the development of more efficient and reliable steelmaking processes.
}
\end{abstract}

\maketitle
\section{Introduction}
\label{sect:intro}
\par
Iron and steel are essential materials in various sectors due to their versatile properties, ranging from construction to automotive to aerospace industries. The production of high-quality steel is a complex process, and one of the major challenges faced during the production is the clogging of submerged entry nozzles (SEN) due to the interaction between the liquid iron (Fe(liq)) and various refractory oxides, including alumina (Al$_{2}$O$_{3}$), calcia (CaO), magnesia (MgO), silica (SiO$_{2}$), and zirconia (ZrO$_{2}$). The clogging phenomena are significantly influenced by the low wettability of these refractory oxides with molten steel, making it a critical factor in controlling continuous casting processes. The wettability is governed mainly by the interfacial interaction between molten steel and the refractory materials, which is influenced by their individual material properties and the operating temperature. Previous studies have focused on experimental characterization and phenomenological models, which, while important, do not provide a full understanding of the molecular mechanisms governing these interactions. 

In this work, we present a first-principles approach to study the wettability of liquid Fe on various refractory oxides at the steelmaking temperature $T = 1823$\,K ($\sim1550\,^{\circ}$C). We combine the density functional theory (DFT) to compute the dielectric responses with a Casimir-Lifshitz model in order to calculate the dispersion forces to determine the interactions across the interfaces of the material systems. 
Motivated by the challenges faced in mitigating SEN clogging, we seek to extend the understanding of the interactions by utilizing the atomistic calculations together with modeling the macroscopic electromagnetic modes by the dispersion forces. The fundamental principles for modeling weak interactions in materials at different length scales are discussed by for instance Fiedler, et al.\cite{Fiedler2023}
Furthermore, we verify the crystalline structures of the oxides by X-ray diffraction, and we measure the contact angles of the Fe(liq)-on-oxide systems by a sessile drop method. The contact angle, a key property in the description of wettability of the substrate, is governed by the cross-interface forces. In the analyses, we determine the related Hamaker constants\cite{hamaker1937london} by 
employing the DFT dielectric response functions in the 
calculations of the Casimir-Lifshitz free energies. This proposed methodology serves as an essential component in understanding the interplay between the liquid and the oxide particles or surfaces. It will enable prediction of wettability at high temperatures as well as offer a deep insight into the interactions at a molecular level. This approach is particularly useful for systems where direct measurement of surface and interfacial energies is challenging, such as Fe(liq)/oxide interfaces at high temperatures. 
\par
By integrating our theoretically determined Hamaker constants with the experimental data of the contact angles, we analyze and compare the wettability of the different refractory oxides by liquid Fe. Since it is more convenient to experimentally explore liquid-oxides interaction at lower temperature, we present also the wettability of a liquid tin-bismuth alloy on the oxides at the temperature 473\,K ($\sim 200\,^{\circ}$C).\cite{Morando2014,Glaser2023} 
Our findings suggest that the interaction energies are crucial in describing the wettability, and the SEN clogging phenomenon remain to a large extent invariant across various scenarios. 
\par
In-depth understanding of the interfacial interactions is critical for further advancements in material processing techniques.
Our investigation of the wettability aims to support fundamental understanding of the submerged entry nozzle clogging phenomena, and we therefore focus on the interaction of liquid iron with the five main refractory oxides and at the specific temperature of steelmaking. 
%
\section{Theoretical details}
\label{sect:theo}
\begin{figure}[h!]
  \centering
  \includegraphics[width=0.98\columnwidth]{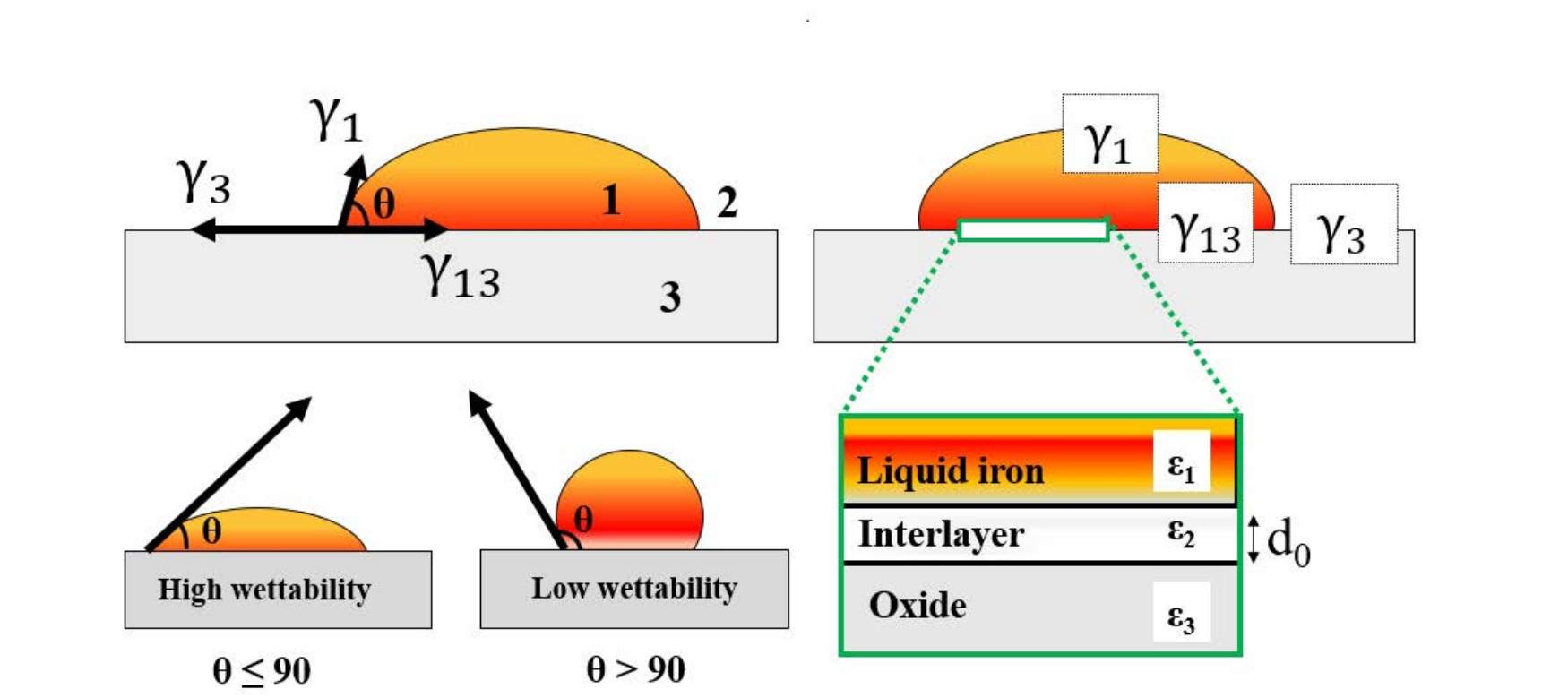}
  \caption{ 
  The work of adhesion $W$ is determined from the interfacial energy $\gamma_{13}$, the surface tension $\gamma_1$ of the liquid iron, and the surface energy $\gamma_3$ of the refractory oxide. 
  The material properties are represented by the dielectric function $\varepsilon_i(\omega)$.
  We model the Casimir-Lifshitz cross-interface interactions with a three-layer system, Fe(liq)//oxide, where the interlayer describes an auxiliary gap with thickness $d_0$ between the liquid iron and the oxide surface. 
  The resulting contact angle $\theta$ is a measure of the wettability.   
  }
\label{fig:wetting}   
\end{figure}

The applied methodology utilizes the permittivity of the material, describing the response to external electromagnetic field, to determine the cross-interface interaction from the change in zero-point energies associated with the surface modes.\cite{Dzyaloshinskii1961} 
We relax the crystalline structure of the materials and compute their complex dielectric functions with the Kohn-Sham approach within the DFT. From the Kramers-Kronig relation, the dielectric functions are evaluated on the imaginary axis for the temperature-dependent Matsubara frequencies. The response functions are thus the basis for computing the Casimir-Lifshitz free energies and the corresponding forces. From that we determine the DFT-derived Hamaker constants, and relate them to the contact angles $\theta$ within Young's theory for ideal surfaces:\cite{young1805}
$\gamma_1 \cos\theta + \gamma_{13} = \gamma_3$, where the gravitational forces are neglected and $\gamma$ is defined by Figure\,\ref{fig:wetting}.
%
Here, the work of adhesion $W$ = $\gamma_1 + \gamma_3 - \gamma_{13}$ is a critical parameter, quantifying the energy associated with the adhesion of two different materials. The work of adhesion can be derived from the Hamaker constant $A_{123}$ by introducing an auxiliary gap distance $d_0$ between the two interacting surfaces:
$W = A_{123}/12 \pi d_0^2$.\cite{Israelachvili2011,Sernelius2001}
With that approach, the interfacial description of the Fe(liq)/oxide system becomes a three-layer model Fe(liq)/interlayer/oxide, where the thickness of the intermediate layer is $d_0$. From our results, we will argue against the notion of a universal value for $d_0$ for all oxides, and instead emphasize that the different characteristics of each system are distinct and important.

%
\subsection{DFT modeling of the materials}
\label{sect:dft}
The atomistic calculations are performed within the DFT framework, employing the projector augmented wave method with the $GW$-type core potentials. 
Given the wide-gap insulating character of the oxides, we employ the generalized gradient approximation that has been revised for solids by Perdew, et al. (PBEsol).\cite{PBEsol2008} We correct the PBEsol band-gap energy with the hybrid functional by Heyd, Scuseria, and Ernzerhof (HSE)\cite{HSE2003} with a 30\% Fock exchange.  
These computations are carried out utilizing the Vienna Ab initio Simulation Package (VASP).\cite{vasp3,vasp1999} 
Spin-orbit coupling is neglected because the spin-split of the oxides is small, it affects the band dispersion only locally, and its contribution to the total energy is marginal. Its impact on the effective carrier masses could be strong,\cite{Persson1997} but that will not be reflected on the overall optical properties nor on the dielectric responses. 
\par
The valence configurations of the elements should include semicore states to include also energy transitions associated with high-energy photons. Therefore, we choose 
Al: $2s^2p^63s^2p^1$ (see also discussion in Ref.~\cite{BostromKuthe2023}),
Ca: $3s^2p^64s^2$,
Fe: $3s^2p^6d^64s^2$,
Mg: $2s^2p^6 3s^2$,
O: $2s^2p^4$,
Si: $2s^2p^63s^2p^2$, and
Zr: $4s^2p^6d^25s^2$.
The compounds are described by their primitive cells, and the irreducible Brillouin zones are typically sampled by $6\times 6 \times 6 $ {\bf k}-meshes. A quasi-Newton variable metric algorithm is utilized for the structural relaxation with a cut-off energy of 800\,eV and to an accuracy of $10^{-4}$\,eV/\AA\ for the forces on each atom. 
Since the optical properties depend on the split-off energies, and the accuracy of determining these energy states depends in turn on bond lengths and bond angles,\cite{Persson1999} we relax the lattice parameters and the atom positions simultaneously. 
Thereafter, the charge density is generated with a 600\,eV cut-off energy, employing Bl\"{o}chl's linear tetrahedron integration for the oxides and the Gaussian smearing technique for iron, and iterated in the electronic self-consistent loop to reach an energy accuracy of $10^{-6}$\,eV. Solid Fe is paramagnetic just below the melting point, crystallized in a body centered cubic (bcc) structure.\cite{lyman1973metallography} Therefore, and to describe the paramagnetic liquid phase of Fe, we use a simplified model of its atomic configuration. That is, we assume bcc structure with a weight density corresponding to liquid Fe at 1823\,K, namely 6.978\,g/cm$^3$. Moreover, the neighboring atoms have opposite spin directions to ensure a perfectly paramagnetic phase.  We justify this model by the fact that the surface energy of bcc Fe below the melting point is very close to the surface tension of liquid Fe at 1823\,K.\cite{schonecker2015thermal} Moreover, one could expect that the ensemble average of the atom coordination and the bond lengths should be bcc-like over time, and therefore also the average dielectric response function of liquid Fe should be reasonably close to that of solid Fe(bcc) at $T \approx 1800 $\,K, though somewhat smeared out. 
We therefore employ a Gaussian smearing of $k_BT = 0.16$\,eV, where $k_B$ is Boltzmann's constant, which is relevant for the considered temperature.
Considering some inexactness in measurements at high temperatures, we show that the resulting response function actually agrees rather accurately with the experimental data.
\par
From the electronic structures of the compounds, the imaginary part of the macroscopic dielectric functions 
$\varepsilon(\omega)$  = $\varepsilon'(\omega)$ + $i\varepsilon''(\omega)$ 
is calculated. With the independent single-electron eigenfunctions, the responses function has contribution from electronic interband transitions, Drude model of intraband transitions for the metals, and ionic vibrations for the oxides. Local field effects are neglected. In the long-wavelength limit, the respective parts read
%
\begin{subequations}
\begin{eqnarray}
\varepsilon^{''}_{{\rm elec}}(\omega) &=&\lim_{{\bf q} \rightarrow 0} \frac{4\pi^2e^2}{\Omega q^2}
 \sum_{v,c,{\bf k}}  \delta (\epsilon_{c,{\bf k}} - \epsilon_{v,{\bf k}} - \hbar \omega) \nonumber \\
&\times&\langle   u_{c,{\bf k}+{\bf e}_{\alpha}q} | u_{v, {\bf k} }   \rangle
\langle   u_{v, {\bf k} } |  u_{c,{\bf k}+{\bf e}_{\alpha}q}   \rangle \,,\\
\varepsilon^{''}_{{\rm Drude}}(\omega) &= &\frac{\Gamma \, \omega_{\rm pl}^2}{\omega(\omega^2 + \Gamma^2)}\, , \ \ 
\mathrm{and} \\
\varepsilon^{''}_{{\rm vibr}}(\omega) &=&  
\sum_j \frac{ S_j\,\omega_{\rm TO,j}^2 \Gamma_j \omega}
{(\omega_{{\rm TO},j}^2 - \omega^2)^2 + (\Gamma_j\omega)^2}\, ,
\end{eqnarray}
\label{eq:eps2c}
\end{subequations}
\ignorespacesafterend
for the three Cartesian directions ${\bf e}_{\alpha}$. 
Here, $\hbar$ is the reduced Planck constant, $\Omega$ is the unit-cell volume and $u_{v/c}$ is the cell periodic part of the valence ($v$) or conduction ($c$) state eigenfunction with the energy $\epsilon_{v/c,{\bf k}}$. For the oxides, the accuracy of especially electronic contribution can depend strongly on the size of the  {\bf k}-point grid,\cite{Crovetto2016} and we therefore use a $12\times 12 \times 12$ {\bf k}-mesh although the values of the low-frequency dielectric constants are  sufficiently converged already for the charge density from the $6\times 6 \times 6$ mesh. The dipole-active longitudinal optical (LO) phonons and the corresponding transverse optical (TO) modes contribute to the dielectric response as vibrations build up an electric field that screens the carriers. $\Gamma_j$ is the damping and $S_j$ is the oscillator strength of the $j$th mode in its vibration direction. From the density functional perturbation theory we compute the Hessian matrix of the ionic displacements to model the optical phonons. 
For metallic Fe, the accuracy in determining the unscreened plasma frequency $\omega_{\rm pl}$, which describes the screening from the nearly free carriers in the partially occupied bands, requires a large $60\times 60 \times 60$ {\bf k}-mesh.
The total imaginary part of the dielectric function is the sum of the contributions: the electronic and Drude for metals, and the electronic and ionic for the oxides. For the average response function, we take the arithmetic mean of the three Cartesian directions.
 
Quantities related to interaction energies can be obtained directly from the imaginary part of the dielectric function. \cite{Dzyaloshinskii1961,NinhamParsegianWeiss1970,Sernelius2001} With the finite temperature formalism in terms of Matsubara frequencies $\omega_n = n\,2 \pi k_B T/\hbar$ (with $n = 0, 1, ...$), the response function is evaluated on the imaginary axis by the Kramers-Kronig relation:\cite{landau1980statistical}
%
\begin{equation}
\varepsilon(i \omega_n) = 1+\frac{2}{\pi}\int_0^\infty d\omega\, \frac{ \omega\, \varepsilon''(\omega)}{\omega^2+\omega_n^2}.
\label{eq:KramKronEq}
\end{equation}
This spectrum is real-valued and decays smoothly towards one at high frequencies $\omega_{n\rightarrow \infty}$.

\subsection{Casimir-Lifshitz dispersion forces}
\label{sect:casimir}
The Casimir-Lifshitz contribution to the interaction free energy between two surfaces across a gap of thickness $d_0$ (see Figure~\ref{fig:wetting}) can formally be calculated by\cite{Dzyaloshinskii1961}  
\begin{equation}
\Delta E_{123} = \frac{k_BT}{2 \pi} {\sum_{n=0}^\infty}{}^\prime \int\limits_0^\infty dq\,q \sum_{\sigma}\ln(1- r_{\sigma}^{21}r_{\sigma}^{23}
 \mathrm e^{-2\kappa_2 d_0}), 
\label{eq:LifFreeEnergy}
\end{equation}  
where $\sigma= $ TE and TM. The prime in the sum indicates that the first term ($n$ = 0) should be weighted by $1/2$, and the upper limit of the summation is about 500 in our calculations. The  Fresnel reflection coefficients between surfaces $i$ and $j$ for the transverse magnetic (TM) and transverse electric (TE) polarizations are given by
\begin{equation}
    r_{\rm TE}^{ij} = \frac{\kappa_i-\kappa_j}{\kappa_i+\kappa_j}\,;  \,\,\,\,\, r_{\rm TM}^{ij} = \frac{\varepsilon_j\kappa_i-\varepsilon_i \kappa_j}{\varepsilon_j \kappa_i+\varepsilon_i \kappa_j} \,. 
    \label{eq:rtTETM}
\end{equation}
Here, $\kappa_i= ({q}^2+\varepsilon_i\omega_n^2/c^2)^{1/2}$ with $i= $1, 2, and 3, and where $c$ is the speed of light. This produces an important contribution to the surface free energies. However, when the two surfaces are very close only the non-retarded transverse magnetic surface modes contribute.  
\par
The work of adhesion is the energy required to separate the two surfaces, which here is direct proportional to the Hamaker constant.
In our analysis, we compute the Hamaker constants by means of the Casimir-Lifshitz interactions in the non-retarded limit and with the variable substitution $\,q \equiv z/d_0$:
%
\begin{eqnarray}
A_{123}&=& -12\pi d_0^2 \, \Delta E_{123} = -6 {k_B T} 
\sum_{n=0}^\infty {}^\prime\,\int_0^\infty dz\,z
 \nonumber \\
&\times&
\ln\Big(1-\frac{(\varepsilon_1-\varepsilon_2) (\varepsilon_3-\varepsilon_2)}{(\varepsilon_1+\varepsilon_2) (\varepsilon_3+\varepsilon_2)}e^{-2 z}\Big)\,,
\label{eq:Hamaker}
\end{eqnarray}
employing the DFT dielectric response functions of the liquid iron ($\varepsilon_1$), ideal air ($\varepsilon_2 = 1$), and the oxides ($\varepsilon_3$). 
We denote this three-layer system with the auxiliary interlayer by Fe(liq)//oxide, i.e. with a double-forward slash symbol.
\par
The thickness of the intermediate layer is obtained by combining Young's equation with the work of adhesion, 
as described by Dupr\'{e},\cite{Dupre1869} yielding 
%
\begin{equation}     
  d_0 = \sqrt{\frac{A_{123}}{12 \pi \gamma_1 (\cos\theta + 1)}}.
\label{eq:d0}
\end{equation}

\section{Experimental details} 
\label{sect:expt}
The primary purpose of the room-temperature X-ray diffraction (RT-XRD) in this study was to validate the lattice constants obtained from DFT relaxation of the crystal structures. The RT-XRD analysis was specifically conducted on refractory oxide powder particles. 
The measurements were performed using a Bruker D2 diffractometer, with a scanning range of 5$^\circ$ to 90$^\circ$, a scan speed of 5$^\circ$ per minute, and a step size of 0.02$^\circ$. The RT-XRD source operated at 40\,kV and 15\,mA, utilizing copper K$\alpha$ radiation at a wavelength of approximately 0.154\,nm.
\par
The contact angles were measured with the support of a high-temperature sessile drop experimental setup (Figure~S1 in Supporting Information). A horizontal Entech electrical tube furnace with Kanthal Super heating elements and an alumina reaction tube (70\,mm inner diameter) was used. On one end of the tube, an internally water-cooled quenching chamber seals the tube with a Viton o-ring, and on the opposite end the tube was sealed with a water-cooled aluminum cap with a sealed borosilicate glass window. The specimen was placed on a graphite carriage that transports the specimen in and out of the hot zone of the furnace, guided on a graphite track. A water-cooled pushrod, sealed with a shaft packing, was connected to the graphite carriage, which in turn is fastened to a screw drive that enables the precise positioning and movement of the specimen for controlled heating and cooling rates. A Eurotherm controller maintains the target furnace temperature with an even temperature zone, defined as $\pm2$\,K from the target temperature. 
A controller thermocouple was a Type B thermocouple (mounted in the wall of the furnace), which was positioned in the hot zone of the furnace outside of the reaction tube. Type C thermocouple with $1\%$ accuracy was mounted axially in the pushrod with the thermocouple tip inside the carriage body positioned directly under the specimen substrate. After pushing the sample holder, the furnace was evacuated and backfilled with pure argon (Ar) at least three times. High-purity Ar gas ($>99.999\%$) was used with a flow rate of 0.1\,liter/minute using a Bronkhorst mass flow meter for this purpose. After refilling, furnace hearing was started on EuroTherm controller at a controlled rate of 2\,K/minute. A Leica V-LUX3 digital camera with full high-definition video ($1920\,\times\,1080$ progressively displayed pixels) capability at 30 frames per second was mounted with a view into the borosilicate glass window on the end of the furnace. Video recordings were initiated when the specimen melts at $\sim 1813$\,K. The melting occurred over the temperature range and was not instantaneous. The images were extracted from the video frames. The contact angle measurement was then analyzed using images with low-bond axisymmetric drop shape analysis,\cite{stalder2010low} which involved fitting the Young-Laplace equation to image data.  Measurements are taken at two critical times: initially when the specimen starts to melt, denoted as $@t_{0}$, capturing the immediate formation and behavior of the droplet, and subsequently at a later stage, referred to as $@t_{c}$, where average values of the contact angles are calculated as the droplet stabilizes.
The contact angles were obtained from images captured during the experiment. In the experiment, the metal transitioned into a molten state within a specified temperature range and stabilized around 1813\,K ($\sim 1540\,^{\circ}$C). The contact angle was initially measured at the melting point at $@t_{0}$ and the average value was subsequently calculated from five images taken between 80--200 minutes post-melting at $@t_{c}$. In this later stage of measurement, the deviation from the average value was at most $2^{\circ}$.
This image-based measurement approach ensured that the reported contact angles were derived from a stable and consistent visual record. The droplet diameter was in the order of $\sim 1$\,cm; see Figure~S2 in Supporting Information for details. 
\par
The contact angles for the tin-bismuth/oxide systems consider a liquid metal alloy with 40\,wt\% bismuth, and were measured in open air rather than in a furnace. Using a Sn-Bi wire sourced from Chip Quik Inc.(Ontario, Canada) the alloy was heated to about 472\,K ($200\,^{\circ}$C) until a droplet formed. This droplet was then placed on the oxide substrate. A camera captured the image of the droplet in situ, and the DropSnake method was later used for image analysis defining the contour of the drop as a versatile B-spline curve.\cite{Stalder2006} The technique analyzed the fitted curvatures to provide measurements of contact angles. The measurements were taken for two specific angles, one when the left side of the droplet contacts the substrate, and another when the right side touches. This helped to understand the droplet's interaction with the oxide surface. Experiments were conducted in open air rather than under Ar gas condition to avoid the complexities associated with setting up a controlled inert atmosphere for low melting-point materials. In this study, stringent oxidation protection was not considered for low melting of this metal alloy. However, open-air conditions simplified the experimental setup and were sufficient for the measurement of the wettability of the tin-bismuth alloy.
%
\section{Results and Discussion} \label{results}
\subsection{Crystal structures of the oxides} 
The crystal structures of the oxides are relaxed and the electronic structures are computed with the PBEsol exchange-correlation functional. The lattice parameters are presented in Table~\ref{tab:DFTrel}. One can observe that the theoretically determined crystal structures agree very well with our RT-XRD analysis (spectra in Supporting Information, Figures S3 to S7) as well with as earlier published experimental data. Deviation of the lattice parameters
between different reported measurements is small for these oxides, also in early publications, and the data are thus well established. We therefore choose to present only two representative values for each lattice parameter: one from the handbook by Samsonov\cite{Samsonov1973} and one from the Landolt-B\"{o}rnstein database.\cite{LandoltBornstein} 
Furthermore, the table presents the calculated $\Gamma$-point direct band-gap energies employing the hybrid functional with 30\% exact exchange in order to generate proper gap values. These energies are compared with available and reasonable experimental data from the two main literature references. One notices a significant deviation of the reported energies between different measurements, which probably can be ascribed to variations in the crystal quality regarding defect inclusion. Hence, values of the band-gap energy are not fully established with high precision. 
%
\begin{table}[h!]
\caption{Lattice parameters $a$, $b$, $c$, and $\beta$ for the seven refractory oxides, determined  with the PBEsol exchange-correlation functional and the RT-XRD analysis. 
The direct band-gap energy $E_{g,\Gamma}^{dir}$ refers to 
the $\Gamma$-point, obtained with the hybrid functional HSE. 
The corresponding experimental energies represent typically the optical gaps.}
\renewcommand{\arraystretch}{1.2}
\setlength{\tabcolsep}{1.5\tabcolsep}
\begin{tabular}{l c c c } 
 \hline  \hline 
    Oxide &  DFT & \hspace{0.1cm}RT-XRD & Literature \\
 \hline 
\multicolumn{2}{l}{Al$_2$O$_3$; R$\overline{3}$c}&&\\
	\hspace{0.3 cm} $a $ [\AA ] & 5.137 & 5.131 &  5.128\,  \cite{Samsonov1973} 5.13\,\cite{LandoltBornstein}    \\
	\hspace{0.3 cm} $\beta $ [$^\circ$] & 55.34 & 55.28 & 55.28\,\cite{Samsonov1973} 55.27\,\cite{LandoltBornstein}   \\
    \hspace{0.3 cm} $E_{g,\Gamma}^{dir}$ [eV] & 8.7&&
    2.5; 3.59\,\cite{Samsonov1973}
    7; 9; 9.26\,\cite{LandoltBornstein}\\
\multicolumn{2}{l}{CaO; Fm$\overline{3}$m} &&\\
	 \hspace{0.3 cm} $a $ [\AA ] & 4.762 &    &  4.799\,\cite{Samsonov1973} 4.810\,\cite{LandoltBornstein} \\
     \hspace{0.3 cm} $E_{g,\Gamma}^{dir}$ [eV] & 6.8&&
     5.59\,\cite{Samsonov1973}
     7.085; 7.7\,\cite{LandoltBornstein}\\
\multicolumn{2}{l}{MgO; Fm$\overline{3}$m}&&\\
	\hspace{0.3 cm} $a $ [\AA ] & 4.211 & 4.2222 & 4.20\,\cite{Samsonov1973}  4.212\,\cite{LandoltBornstein}\\
    \hspace{0.3 cm} $E_{g,\Gamma}^{dir}$ [eV] & 6.9&& 
    7.29\,\cite{Samsonov1973}
    5.13; 7.833,\cite{LandoltBornstein}\\
\multicolumn{2}{l}{SiO$_2$; P$3_1$21}&&\\
    \hspace{0.3 cm} $a $ [\AA ] & 4.905 & 4.9191& 4.913\,\cite{Samsonov1973} 4.9134\,\cite{LandoltBornstein} \\
	\hspace{0.3 cm} $c $ [\AA ] & 5.405 & 5.4105& 5.405\,\cite{Samsonov1973}  5.4046\,\cite{LandoltBornstein} \\
     \hspace{0.3 cm} $E_{g,\Gamma}^{dir}$ [eV] & 8.5& & 9.1; 9.43\,\cite{LandoltBornstein} \\
\multicolumn{2}{l}{SiO$_2$; Fd$\overline{3}$m}&&\\
    \hspace{0.3 cm} $a $ [\AA ] & 7.360 & & 7.1487\,\cite{Wright1975} 
    7.1393\,\cite{LandoltBornstein}\\
     \hspace{0.3 cm} $E_{g,\Gamma}^{dir}$ [eV] & 7.7 & &  \\     
\multicolumn{2}{l}{ZrO$_2$; P2$_1$/c}&&\\
    \hspace{0.3 cm} $a $ [\AA ] & 5.126 & 5.1487 & 5.17\,\cite{Samsonov1973} 5.127\,\cite{LandoltBornstein} \\
    \hspace{0.3 cm} $b $ [\AA ] & 5.213 & 5.2098 & 5.26\,\cite{Samsonov1973} 5.202\,\cite{LandoltBornstein}\\
    \hspace{0.3 cm} $c $ [\AA ] & 5.292 & 5.3149 & 5.30\,\cite{Samsonov1973} 5.31\,\cite{LandoltBornstein}\\
   \hspace{0.3 cm} $\beta $ [$^\circ$] & {99.57} & 99.204&  99.90\,\cite{Samsonov1973}  99.18\,\cite{LandoltBornstein}\\
   \hspace{0.3 cm} $E_{g,\Gamma}^{dir}$ [eV] & 6.0&&
   2.0\,\cite{Samsonov1973}
   5.83; 7.09\,\cite{LandoltBornstein}\\
\multicolumn{2}{l}{ZrO$_2$; P4$_2$/nmc}&&\\
    \hspace{0.3 cm} $a $ [\AA ] & 3.584 &    & 3.578\,\cite{LandoltBornstein}  \\
    \hspace{0.3 cm} $c $ [\AA ] & 5.175 &    & 5.19\,\cite{LandoltBornstein}  \\
   \hspace{0.3 cm} $E_{g,\Gamma}^{dir}$ [eV] & 6.1& &  0.74; 3.3\,\cite{LandoltBornstein}\\
 \hline  \hline  
\end{tabular}  
\label{tab:DFTrel}
\end{table}
\par
Alumina is the main refractory component in the nozzle walls for steelmaking processes. The compound is thermodynamically stable in the $\alpha$-phase, which is a rhombohedral structure (space group \#167; R$\overline{3}$c, with $D_{3d}$ as the underlying point group). The RT-XRD determined lattice parameters of Al$_2$O$_3$ confirm good crystalline quality. The compound is often presented based on the three times larger hexagonal unit cell, and the RT-XRD data are then $a = 4.761$\,\AA\ and $c/a = 2.730$. The corresponding DFT values are $a = 4.771$\,\AA\ and $c/a = 2.726$, yielding a unit cell volume of $\Omega = 8.55$\,\AA$^3$/atom. There are the two types of bonds: each Al atom has three bonds with the length of $\delta_b$(Al--O) = $1.86$\,\AA\ and three with the length of $\delta_b$(Al--O) = $1.97$\,\AA. According to the empirically measured covalent atomic radii by Slater,\cite{Slater1964} the Al atom has a radius of $1.25$\,\AA\ and the O atom has radius $0.60$\,\AA, thus one could expect a smallest bond length in the order of $1.85$\,\AA.
The band-gap energy of 8.7\,eV from the DFT calculations is in line with the recent measurements,\cite{LandoltBornstein} while older experiments underestimate the gap quite significantly.\cite{Samsonov1973} 
\par
At room temperature, and also at 1823\,K, calcia crystallizes in a face-centered cubic ('cub') phase, more precisely the rock salt structure (space group \#225; Fm$\overline{3}$m with point group $O_{h}$). The covalent radius of the Ca atom is the largest of all elements for the considered oxides, e.g. $\sim 44$\% larger than that of Al, and therefore the unit cell volume is large: $\Omega = 13.50$\,\AA$^3$/atom. Each atom in CaO is six-fold coordinated with a bond length of $\delta_b$(Ca--O) = $2.38$\,\AA. The calculated band-gap energy at the $\Gamma$-point is $6.8$\,eV, which is in the same order as the other refractory oxides.
However, our RT-XRD analysis of CaO powder does not support single crystalline cubic CaO. Instead, the powder is composed by a mixture of phases with a strong weight fraction of hexagonal calcium hydroxide. That is not surprising as calcia is exothermally hygroscopic under ambient humidity condition.\cite{Whitman1926} The experimentally determined lattice parameters are $a = 3.592$\,\AA\ and $c/a = 1.371$, which are consistent with our theoretical data for trigonal Ca(OH)$_2$ with space group P$\overline{3}$m1: $3.555$\,\AA\ and $1.334$, respectively. The powder contains also of calcium carbonate, CaCO$_3$, which exists in various phases with formation energy $<0.02$\,eV/atom above hull.\cite{MaterialsProj2013}. The most stable polymorph at room temperature is calcite, which has a similar structure as Al$_2$O$_3$. DFT yields $a = 6.308$\,\AA\ and $\beta = 46.57^\circ$, which agree with data from several reports.\cite{LandoltBornstein} 
The RT-XRD detects however a mixture of an orthorhombic and a monoclinic structure, which suggests inclusion of aragonite. In addition, the RT-XRD reveals a hexagonal phase of CaO; see Figure~S3 in Supporting Information.
The structure has, to our knowledge, not been experimentally observed before. The lattice constants are $a = 3.945$\,\AA\ and $c/a = 1.158$. 
A DFT calculation of hexagonal CaO (\#194, P6$_3$/mmc) seems to support that structure with $a = 3.952$\,\AA\ and $c/a = 1.199$, despite a significantly larger lattice constant in the [0001] direction. The DFT formation enthalpy of this hexagonal phase is however $\sim 90$\,meV/atom higher than that of its cubic counterpart, indicating that the polymorph is unlikely to be formed as a single-crystal high-quality substrate.
\par
Magnesia is thermodynamically stable in a rock salt structure similar to that of cubic calcia. Also the $\Gamma$-point gap energies of the two cubic compounds are very similar: $6.9$\,eV for MgO. 
However, since the Mg atom is smaller than Ca, the MgO cell volume and bond length are smaller than for CaO:  $\Omega = 9.33$\,\AA$^3$/atom and $\delta_b$(Mg--O) = $2.11$\,\AA. Actually, the volume and bond length are closer to those of Al$_2$O$_3$. Both Mg and Al are  period-3 elements with rather similar atomic radii, though Mg is about $\sim 20$\% larger. The obvious difference between MgO and Al$_2$O$_3$ is the cation valence configurations (group-IIA versus group-IIIA), which thereby form different covalent bond symmetries by the octet rule.
\par
Silica is composed of the two most abundant elements in the Earth's crust, and the quartz family of silica exists in different varieties. At room temperature, $\alpha$-SiO$_2$ is the most stable phase, whose structure is trigonal ('tri'; \#152; P$3_1$21 with point group $D_{3}$). The cell volume is $12.51$\,\AA$^3$/atom, and each Si has four bonds with $\delta_b$(Si--O) = $1.62$\,\AA. At higher temperatures, this SiO$_2$(tri) phase is easily and abruptly transformed to other phases, and at 1823\,K it is expected to become face-centered cubic $\beta$-cristobalite. The computed formation energy of ideal $\beta$-cristobalite (\#227; Fd$\overline{3}$m based on $O_h$) is $\sim 19$\,meV/atom above the convex hull. The unit cell volume of this cubic phase ($16.61$\,\AA$^3$/atom) is notably larger than that of the trigonal counterpart, but this fact is not reflected in the bond distance ($1.59$\,\AA), nor is it explained by any change in the coordination number. 
Experimental analysis has identified the high-temperature compound as a Fd$\overline{3}$m structure, however with partially occupied Wyckoff sites.\cite{Wright1975} An alternative phase is a distorted structure that lowers the symmetry to cubic P2$_1$3.\cite{LandoltBornstein} One should therefore not expect perfect agreement between computed and measured lattice parameters. The calculated band-gap energy $8.5$\,eV of SiO$_2$(tri) agrees rather well with the experimental finding that often is in the range of 8.5 to 9.5\,eV. SiO$_2$(cub) has $\sim 1$\,eV smaller gap.
\par
Zirconia is often listed as a high-$\kappa$ dielectric material. The compound crystalizes in a monoclinic structure at room temperature ('mono'; \#14; P2$_1$/c with point group $C_{2h}$). The volume of the unit cell is $\Omega = 11.62$\,\AA$^3$/atom. Due to low crystalline symmetry, each Zn has different bond distances to its surrounding O atoms: $2.05$, $2.06$, $2.15$, $2.15$, $2.17$, $2.23$, and $2.25$\,\AA.
The high-temperature phase of ZrO$_2$ is however tetragonal ('tetra'; \#137; P4$_2$/nmc with $D_{4h}$). The DFT formation enthalpy of ZrO$_2$(tetra), neglecting temperature effects, is $\sim 26$\,meV larger than that of ZrO$_2$(mono). The two phases have very similar band-gap energies: 6.1 versus 6.0\,eV.
Also, the tetragonal cell volume $\Omega = 11.08$\,\AA$^3$/atom is similar to the volume of the monoclinic cell. However, the local bond arrangements are different: in ZrO$_2$(tetra) each Zr atom has four bonds with $\delta_b$(Zr--O) = $2.07$\,\AA. and four with $\delta_b$(Zr--O) = $2.36$\,\AA. From the covalent atomic radii\cite{Slater1964} one would expect that the bond distances should be around $2.15$\,\AA. 

\subsection{Dielectric response functions}   
The frequency-dependent dielectric response function, or the relative permittivity, is a fundamental property to describe the electric displacement field resulting from an applied electric field. It is also the main optical characteristic of a material, represented often by its dielectric constant(s).
Having established the crystalline structures of the oxides, the complex dielectric functions are calculated according to Eq.~\ref{eq:eps2c}. The resulting static $\varepsilon_0$ and high-frequency $\varepsilon_\infty$ dielectric constants (Table~\ref{tab:eps}) are determined at zero frequency by respectively including and disregarding the contribution from the optical vibrations. That approach is somewhat different from the experimental determination. The static dielectric constant is typically measured at very low frequency, so that should be comparable to the DFT values. The high-frequency dielectric constant is however experimentally resolved at a frequency corresponding to an energy that should be sufficiently below the optical gap. Since the dielectric function in that region is increasing as function of the frequency, one could expect a somewhat larger measured value compared to the theory, however, the deviation is typically not large; often between 0.1 and 0.3. 
%
\begin{table}[h]
\caption{The static $\varepsilon_0$ and high-frequency $\varepsilon_\infty$  dielectric constants
of the oxides and in the three Cartesian directions, i.e., the diagonal components of the $3 \times 3$ tensor.
}
\renewcommand{\arraystretch}{1.1}
\setlength{\tabcolsep}{2.2\tabcolsep}
\begin{tabular}{l ccccc} 
 \hline  \hline 
    Oxide & \multicolumn{3}{c}{DFT}          & Expt. \\
    &  $xx$  & $yy$ &   $zz$  &   $xx / zz$   \\
 \hline 
\multicolumn{2}{l}{Al$_2$O$_3$}&& \\
\hspace{0.3 cm} $\varepsilon_0$       & 9.3&$xx$&11.4&
10.5--12; 12.3\cite{Samsonov1973}
9.34/11.54\cite{LandoltBornstein} \\
\hspace{0.3 cm} $\varepsilon_\infty$  & 3.0&$xx$& 3.0&    \\
\multicolumn{2}{l}{CaO}&& \\
\hspace{0.3 cm} $\varepsilon_0$       &15.3&$xx$  &$xx$ &  
3.00--0.10; 11.8~\cite{Samsonov1973}
11.96~\cite{LandoltBornstein} \\
\hspace{0.3 cm} $\varepsilon_\infty$  & 3.4&  $xx$&  $xx$  & 
3.33; 3.385~\cite{LandoltBornstein}\\
\multicolumn{2}{l}{MgO}&&\\
\hspace{0.3 cm} $\varepsilon_0$       & 9.9 & $xx$ & $xx$&  
3.2--3.1; 8--10.5~\cite{Samsonov1973}
9.95~\cite{LandoltBornstein}   \\
\hspace{0.3 cm} $\varepsilon_\infty$  & 2.9 & $xx$ & $xx$&     \\
\multicolumn{2}{l}{SiO$_2$(tri)}&&\\
\hspace{0.3 cm} $\varepsilon_0$      & 4.5 & $xx$ & 4.7 &
3.5--4.1;4.34/4.27\cite{Samsonov1973}
4.42/4.60~\cite{LandoltBornstein} \\
\hspace{0.3 cm} $\varepsilon_\infty$ & 2.4 & $xx$ & 2.4 &  \\
\multicolumn{2}{l}{SiO$_2$(cub)}&&\\
\hspace{0.3 cm} $\varepsilon_0$      & 3.8 & $xx$ & $xx$ & \\
\hspace{0.3 cm} $\varepsilon_\infty$ & 2.0 & $xx$ & $xx$ &  \\
\multicolumn{2}{l}{ZrO$_2$(mono)}&&\\
\hspace{0.3 cm} $\varepsilon_0$      &20.5&22.4&16.2&  
12.5; 23~\cite{LandoltBornstein}\\
\hspace{0.3 cm} $\varepsilon_\infty$ & 4.6& 4.6& 4.3&   \\
\multicolumn{2}{l}{ZrO$_2$(tetra)}&&\\
\hspace{0.3 cm} $\varepsilon_0$      &63.5&$xx$&18.9& \\
\hspace{0.3 cm} $\varepsilon_\infty$ & 4.9& $xx$& 4.5&  \\
 \hline  \hline  
\end{tabular} 
\label{tab:eps}
\end{table}
\par
From Table~\ref{tab:eps} one can observe that all the considered oxides exhibit high-frequency dielectric constants between 2.0 and 4.9, where SiO$_2$(cub) has the lowest and ZrO$_2$(tetra) has the highest value. The spatial anisotropy of the high-frequency dielectric constants is small, at most $\sim 7$\% (for ZrO$_2$) and zero by symmetry reason for the cubic phases. Moreover, the two phases of SiO$_2$ have similar constants, and that is true also for the two phases of ZrO$_2$. The static dielectric constants are however more different between the oxides. These values are also more anisotropic, as it depends on the local bond symmetries. Here, the contribution from the optical phonon vibration is only $\sim$ 2 for the two SiO$_2$ phases, whereas it is $\sim 6$--$8$ for Al$_2$O$_3$ and MgO, $\sim 12$ for CaO, and as much as $\sim 12$--$60$ for ZrO$_2$. The theoretical data agree with experimental results for the most common and well-studied oxides, i.e. Al$_2$O$_3$, MgO, and SiO$_2$(tri), while the data for high-$\kappa$ ZrO$_2$(mono) seems to not be fully established. One can notice that the tetragonal and monoclinic phases of ZrO$_2$ have similar high-frequency dielectric constants, but at low frequencies their dielectric functions diverge by as much as $\sim 3$ times in the $xy$-plane. Notably is also the high spatial anisotropy of the static dielectric constant for the high-temperature phase ZrO$_2$(tetra).
%
\begin{figure}[h!]
  \centering
  \includegraphics[width=0.95\columnwidth]{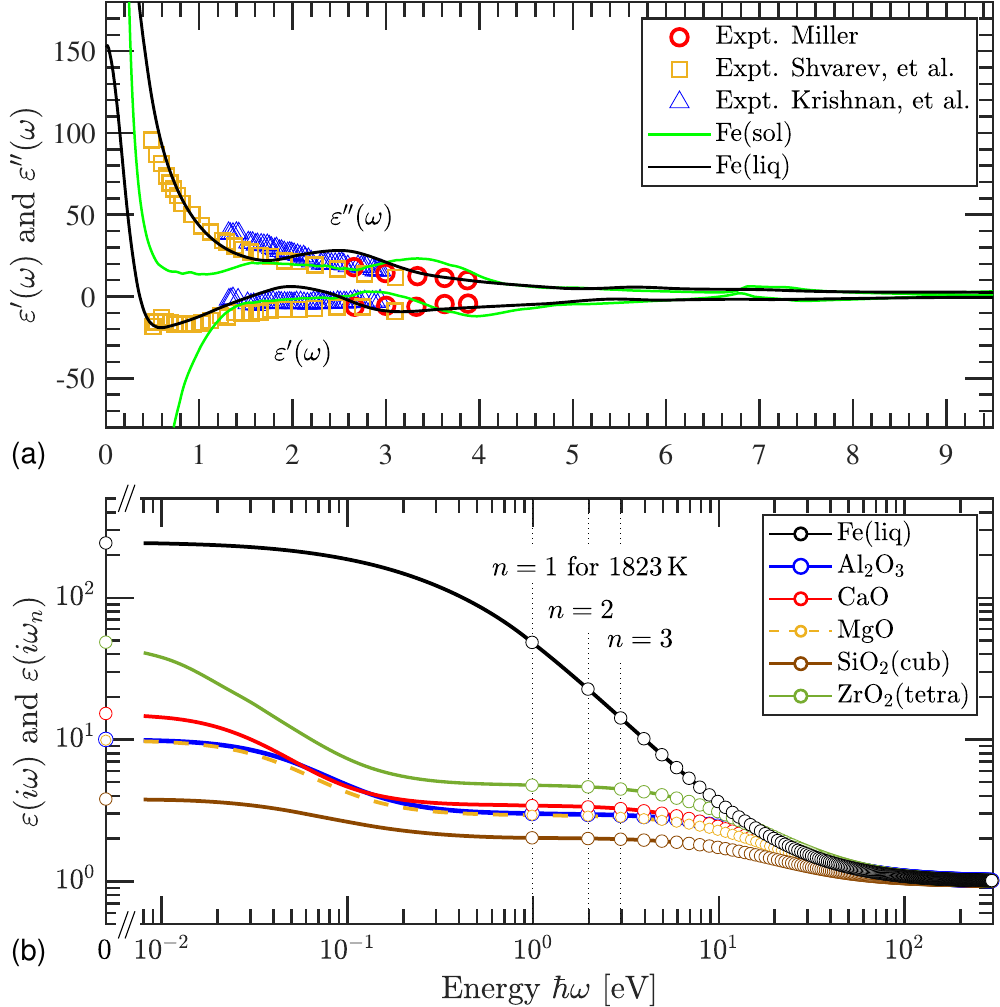}
  \caption{ 
  (a)  The dielectric function of the modeled liquid Fe (Fe(liq); black lines), compared with the solid Fe at room temperature (Fe(sol); green lines) and available experimental data (marks) from Refs. \cite{miller1969optical,shvarev1978effect,krishnan1997optical} 
  (b)  Dielectric functions of liquid Fe and the refractory oxides  elaborate on the imaginary axis. Here, the circles indicate data at the Matsubara frequencies $\omega_{n}$.
  At zero frequency (i.e., $\omega = \omega_{0} = 0$), the data equals $\sim$244 for Fe(liq) and the average static dielectric constants $\varepsilon_0$ for the oxides, which can be obtained also from Table~\ref{tab:eps}.} 
\label{fig:dft_diel}
\end{figure}
\par
The complex dielectric function of liquid iron, Fe(liq), is modeled for the temperature 1823\,K and with an ideal crystalline structure. We verify that the model is reasonable by comparing the resulting spectrum with corresponding spectrum for room temperature solid iron, Fe(sol), as well as by comparing it with available measured dielectric responses; see Figure~\ref{fig:dft_diel}(a).
First, there are clear differences between the dielectric functions of Fe(sol) and Fe(liq). The reason is due to the change in volume, where Fe(liq) has 11\% lower density, but also due to different magnetic configurations. Here, our calculated ferromagnetic Fe(sol) has an intrinsic magnetic moment of 2.6\,$\mu_B$/atom, which is close to measured $\sim 2.4$\,$\mu_B$/atom, whereas Fe(liq) is paramagnetic. That affects the dielectric dispersion especially for energies below 1.5\,eV, but also in the region between about 2 and 4\,eV.
Second, the dielectric function of Fe(liq) is compared with experimental reports by Miller,\cite{miller1969optical} Shvarev et al.,\cite{shvarev1978effect} and Krishnan et al.\cite{krishnan1997optical} The three measured spectra are complementary and overlapping, covering an energy region from about $0.5$ to $4.0$\,eV. In this region our modeled dielectric function of Fe(liq) agrees fairly well with the measured data. Moreover, the theoretical response function goes asymptotically correctly to one at high energies. In addition, based on the dielectric function and using a similar model as Eq.~\ref{eq:LifFreeEnergy}, our calculated surface tension of Fe(liq) is $\gamma_1 = 1.843$\,J/m$^2$, which agrees well with the measured data between $1.83$ and $1.92$\,J/m$^2$ with the adopted mean values of $1.85$, $1.88$, or $1.90$\,J/m$^2$.\cite{BJKeene1993,KCMills2013}
We therefore anticipate that our model for describing the dielectric response of liquid iron is reasonable for exploring the Casimir-Lifshitz dispersion forces of the Fe(liq)/oxide systems.   
\par
The dielectric response functions are elaborated on the imaginary axis according to Eq.~\ref{eq:KramKronEq}, and the resulting spectra for Fe(liq) and the high-temperature phases of the oxides are presented in Figure~\ref{fig:dft_diel}(b). The circles represent the data for Matsubara's frequencies for $T = 1823$\,K. The liquid metal has a much larger dielectric function than the oxides at low frequencies due to the Drude contribution from the (almost) free electrons. On the other hand, the metal does not exhibit the oxides' declining dispersion at $\sim 0.01$--$0.5$\,eV, which is a consequence of that the lattice dynamics does not play an important role at energies higher than $\sim 1$\,eV. 
The first Matsubara frequency (i.e., $n = 0$) yields the static dielectric constant $\varepsilon_0$. For the second Matsubara frequency (i.e., $n = 1$), the oxides' dielectric functions are close to their respective high-frequency dielectric constants $\varepsilon_\infty$. In the energy region up to some tens of eV, the spectrum of ZrO$_2$(tetra) is the largest of the considered oxides, the spectrum of SiO$_2$(cub) is the smallest, whereas  Al$_2$O$_3$, MgO, and CaO have rather similar spectra. The spectra decline smoothly even further at energies that correspond to the respective band-gap energies. At very high energies, the spectra of the oxides go, just like that of Fe(liq), asymptotically to one.   
\par
The dielectric response functions of the oxides and liquids are essential representations for describing the interaction under high-temperature conditions, which in turn also influence material factors such as thermal stability and wettability of the substrates by the liquids.
%
%
\subsection{Wettability of liquid/oxide} 
\label{sect:Fe/oxides}
The strength of the cross-interface interaction in a liquid/substrate system is described by the work of adhesion $W$, which is often represented by the Hamaker constant $A_{123}$. We determine theoretically the Hamaker constants of the Fe(liq)//oxide systems directly from the dielectric response functions according to Eq.~\ref{eq:Hamaker} and at the temperature $T = 1823$\,K. 
In Table~3, 
it is clear that ZrO$_2$ has the largest Hamaker constant of these refractory oxides. The constant of SiO$_2$ is the smallest, whereas Al$_2$O$_3$, MgO, and CaO have rather similar Hamaker constants. That is consistent with the strengths of the oxides' dielectric response functions [Figure~\ref{fig:dft_diel}(b)].
Representing the liquid-substrate interaction only by the Hamaker constant is however not unambiguous. It is a parameter useful for describing, for example, forces between macroscopic objects at sufficiently large  distances, but the three-layer model in Eq.~\ref{eq:LifFreeEnergy} is not applicable for an infinitely small interlayer thickness. 
The reason is that the physical quantity, i.e. the work of adhesion, is proportional to $A_{123}/d_0^2$ and thus involves a finite size of the auxiliary gap. As a consequence, an incorrect determination of the Hamaker constant could generate a correct value of $W$ if also the size of $d_0$ is adjusted. However, that Hamaker constant could not be utilized as a parameter to describe forces as function of the distance. Hence, a consistent determination of $d_0$ is required, accompanying the presentation of $A_{123}$ in order to properly describe the wettability.  
\par
\begin{figure*}[!ht]
  \centering
  \includegraphics[width=1.6\columnwidth]{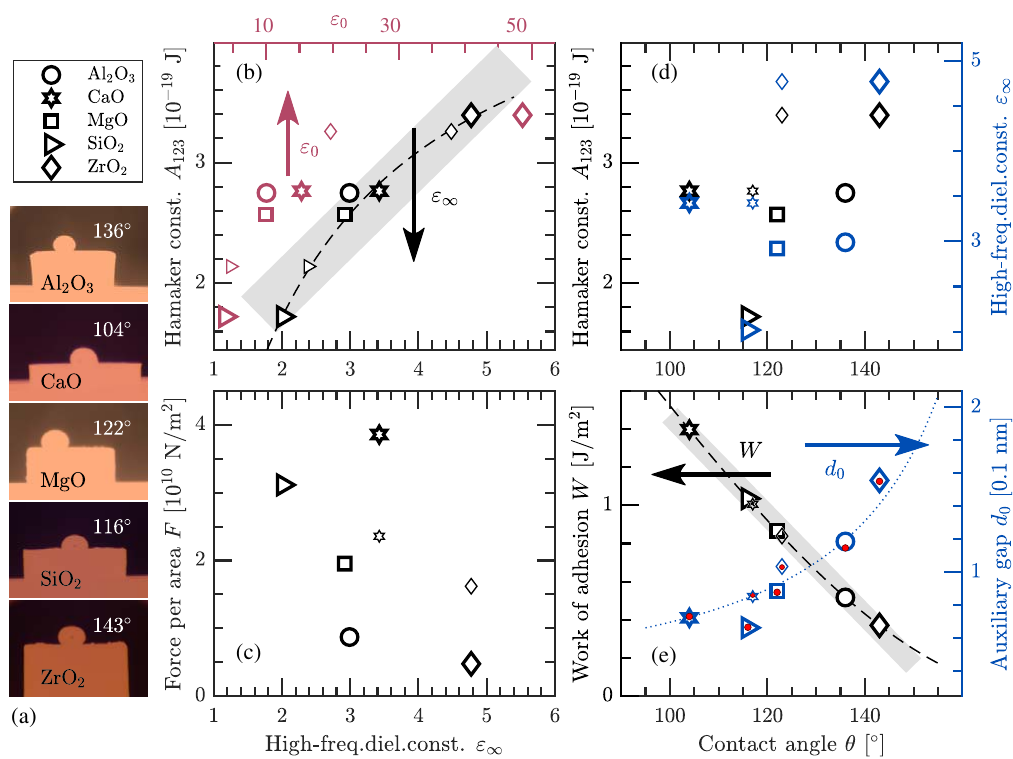}
  \caption{(a) Experimentally determined contact angles of the Fe(liq)-on-oxides; all at $T = $ 1823\,K and under argon flow condition. Subfigures (b) to (d) describe property relations for systems with Al$_2$O$_3$ (circles), CaO (hexagrams), MgO (squares), SiO$_2$ (triangles), and ZrO$_2$ (diamond symbols). Dashed and dotted lines represent fitted curvatures, and gray areas are guides for the eye to indicate an almost linear relation.  
  The smaller marks in subfigure (b) are the corresponding results for the room-temperature phases of SiO$_2$ and ZrO$_2$, whereas the smaller marks in (c), (d), and (e) are the results for CaO and ZrO$_2$(tetra) when using the average values of their contact angles $\theta_{ave}$, i.e. including earlier published data.
  (b) Hamaker constant $A_{123}$ versus the high-frequency  $\varepsilon_\infty$ (bottom axis) and static $\varepsilon_0$ (top axis) dielectric constant. 
  (c) Force $F$ per area versus the high-frequency dielectric constant. 
  (d) Hamaker constant (left axis) and high-frequency dielectric constant (right axis) versus the contact angle $\theta$.
  (e) The work of adhesion $W$ (left axis) and the interlayer gap distance $d_0$ (right axis)  as functions of the contact angle. The red filled circles represent an approximated expression, utilizing the fit of the Hamaker constant in subfigure (b) as generated by Eq.~\ref{eq:A123vseps}.
  }
\label{fig:results}
\end{figure*}

In the experimental investigation of the wettability of melted iron on the different oxide surfaces, a distinct variation in the contact angles is observed from the sessile drop measurements
(Figure~\ref{fig:results}(a) and Table~\ref{table:HamakerFe}. 
The results for the five Fe(liq)/oxides, a direct indicator of their wettability at $T = 1823$\,K, range from moderate wettability to low-wetting surfaces. Calcia presents a contact angle of $104^{\circ}$, suggesting a moderate degree of wettability. A slightly higher contact angle of $116^{\circ}$ is noted for silica, indicating a reduced wettability. This trend of increasing contact angles continues with magnesia, which exhibits a contact angle of $122^{\circ}$, and alumina that shows a further increased contact angle of 136$^{\circ}$, implying an even lower wettability. 
The most significant resistance to wetting is demonstrated by zirconia, with a contact angle of initially $152^{\circ}$ which changes over time to be stabilized at $143^{\circ}$.  For water, a surface is described as superhydrophobic if the water/substrate system has a static contact angle that exceeds $150^\circ$.\cite{SWang2007} Our measured contact angle for Fe(liq)/ZrO$_2$ is close to that very low degree of  wettability, however earlier reported data indicate a smaller angle of around $120^\circ$.\cite{Nakashima1992,Ogino1973} \\
\par
%
%
\begin{table*}[!ht]
\centering
 \setlength{\tabcolsep}{\tabcolsep}
\caption{ Hamaker constants $A_{123}$, interfacial distances $d_0$, works of adhesion $W$, and corresponding forces $F$ per unit area for the Fe(liq)-on-oxide systems. The measured contact angles $\theta$ are presented at the initial time when reaching $T$ = 1823\,K (column denoted $@t_0$), and when measuring at the later stage (denoted $@t_c$); the latter is our confirmed average value of the contact angles with an error bar of $\pm 2^{\circ}$. Cubic SiO$_2$ and tetragonal ZrO$_2$ are expected to be the respective high-temperature phases. The results are compared with experimental data in the literature.}
\renewcommand{\arraystretch}{1.2}
\begin{tabular}{lccccccc}
   \hline  \hline 
 Oxide & $A_{123}$ [$10^{-19}$\,J]\hspace{0.2cm} &  \multicolumn{3}{c}{$\theta$ [arc degree]}  &$d_0$\,[nm] & $W$ [J/m$^2$]& $F$ [$10^9$ N/m$^2$]\\
  \cline{3-5} 
 & &  $@t_{0}$ & $@t_c$ &  Literature & & & \\
   \hline
  Al$_{2}$O$_{3}$ & 2.75 & 134 & 136 & 
  \phantom{00}
  135; l44\cite{Nakashima1992}$^{*}$
  126.9\cite{Shen2019}
  132\cite{kapilashrami2003}
  139; 141\cite{Samsonov1973}
    118--129\cite{Ogino1973}$^{*}$
  & 0.119& 0.52 & \phantom{0}8.71\\
  CaO & 2.76 & 106& 104 & 
   121\cite{Nakashima1992}
   132\cite{Samsonov1973}
   104--114\cite{Ogino1973}$^{*}$
   & 0.072 & 1.40 & 38.56\\
  MgO & 2.57 & 124& 122 &
  128\cite{Nakashima1992}
  133.5\cite{Shen2019}
  123; 130\cite{Samsonov1973}
    \phantom{0}96--122\cite{Ogino1973}$^{*}$
  & 0.089 & 0.87 &  19.53\\
  SiO$_{2}$(tri) & 2.14 & \multirow{2}{*}{117}& \multirow{2}{*}{116} &
  \multirow{2}{*}{110\cite{Nakashima1992}
  105--135\cite{kapilashrami2003b} 
  108\cite{Samsonov1973}
  116--128\cite{Ogino1973}$^{*}$}
  & 0.074 & 1.03 & 27.95\\
  SiO$_{2}$(cub) & 1.72 & &  & 
  & 0.066 & 1.03 & 31.14\\
  ZrO$_{2}$(mono) & 3.26 & \multirow{2}{*}{152}& \multirow{2}{*}{143} & 
  \multirow{2}{*}{119;122\cite{Nakashima1992}
  92; 102; 111\cite{Samsonov1973}
  105--120\cite{Ogino1973}$^{*}$}
  & 0.153 & 0.37 & \phantom{0}4.86\\
  ZrO$_{2}$(tetra) & 3.39 & &  &  & 0.156 & 0.37 & \phantom{0}4.77\\
  \hline \hline
  \multicolumn{5}{l}{$^*$Experimental data at $T$= 1873\,K}
\label{table:HamakerFe}
\end{tabular}
\end{table*}
%
\par
Experimental approaches for measuring contact angles at high temperatures, e.g. $1823$\,K, often overlook the nuanced significance of accurately considering and treating competing phases at the liquid-oxide interface. 
The potential formation of intermediate phases, such as FeAl$_{2}$O$_{4}$ in the Fe(liq)/Al$_{2}$O$_{3}$ system, could complicate a precise measurement of the contact angles. 
This uncertainty arises from the ambiguity in determining the thermodynamic stability of an ideal interface contrary chemical reaction or adsorption, that questions whether the work of adhesion with its auxiliary gap relates to the Fe(liq)/Al$_{2}$O$_{3}$ interface or a Fe(liq)/FeAl$_{2}$O$_{4}$/Al$_{2}$O$_{3}$ interface. However, an ultrathin modified surface layer is not expected to have a decisive significance for the cross-interface interactions.  
Historical sessile drop experiments have reported varying contact angles for Fe/Al$_{2}$O$_{3}$,\cite{Samsonov1973,Ogino1973} perhaps indicating that an intermediate phase of a thicker size influences the measurements. 
However, our result $\theta = 136^\circ$ for alumina is consistent with the most recent data in the literature.\cite{Nakashima1992,Shen2019,kapilashrami2003}
The complexity with competing phases extends to other systems as well. 
For instance, calcia is hygroscopic and Ca(OH)$_2$ is easily formed. That should however be less problematic for the measurement of Fe(liq)/CaO, because H$_2$O is expected to evaporate above $T \approx 785$\,K ($512\,^\circ$C) and the substrate will transform to CaO. Similarly, contamination of CaCO$_3$ will decompose into CaO and CO$_2$ beyond its calcination temperature $\sim 1023$\,K ($750\,^\circ$C). Our measured contact angle of calcia, $\theta = 104^\circ$, is however smaller than earlier published data, though those data span a quite large region of angles.   
It might be that a thin film of liquid calcite is present at the Fe(liq)/CaO interface due to the cross-interface pressure of almost $40$\,GPa; 
such effect is supported by phase diagram from Ivanov, et al.\cite{Ivanov2002}
A perhaps more plausible explanation is that carbon from the CO$_2$ vapor contaminates the molten iron with the release of O$_2$. Support for such effect is that the wettability of alumina is higher by carbon-containing molten iron than by pure liquid iron.\cite{Samsonov1973}
The CaO samples used in our experiment may differ from those referred to in the literature, especially in terms of purity, density, and crystalline structure. 
A related effect to consider is the possibility of oxygen adsorption on the surface of the metal droplet.\cite{kapilashrami2003} 
If the concentration of oxygen at the surface is high, there could be diffusion of oxygen from the surface to the bulk. Such an oxidation process may contribute to the change of the contact angle with respect to time.
Further, in the Fe(liq)/MgO system, the high-temperature interactions could lead to the formation of phases 
like MgFe$_{2}$O$_{4}$, which would alter the interfacial characteristics and, consequently, the contact angles. 
Similarly, in the Fe(liq)/SiO$_{2}$ system, the presence of iron silicates at the interface may impact the wettability and the work of adhesion. 
The Fe(liq)/ZrO$_{2}$ system also presents similar challenges, and in addition, the substrate should be able to transform from monoclinic to tetragonal phase as the substrate is heating up above $1443$\,K ($1170\,^\circ$C). That could be a possible reason for the notably decrease in the contact angle over time. The presence of impurities and increased surface roughness may increase the contact angles by restricting the spread of liquid metal across the ZrO$_{2}$ substrate.
Another issue is associated with the measurement conditions. The choice of applied gas in the furnace, in our case argon, as well as its pressure will affect especially the liquid surface tension. For low and moderate gas pressure, the effect should be neglectable. 
At high applied gas pressure at high temperatures one could expect some further amorphorization of SiO$_{2}$ and an inclusion of the $\beta$-quartz phase.
\par
Overall, however, our measured contact angles of the Fe(liq)/oxide systems agree fairly well within the range of published data in the literature. Nonetheless, for Fe(liq)/CaO and Fe(liq)/ZrO$_{2}$ we will consider also the average value of their contact angles, including the earlier reasonable experimental data: 
$\theta_{ave} = 117^\circ$ and $123^\circ$, respectively.
\par
The Casimir-Lifshitz free energy and the corresponding Hamaker constant depend explicitly on the dielectric response functions of the layers in the liquid//substrate systems. In Figure~\ref{fig:results}(b), we present the Hamaker constants of the Fe(liq)//oxides versus the high-frequency $\varepsilon_\infty$ (bottom axis) and static $\varepsilon_0$ (top axis) dielectric constants of the oxides.   
The Hamaker constant increases almost linearly with the size of the high-frequency dielectric constant. A simple linear regression analysis yields $(0.820 + 0.555\varepsilon_\infty)\times 10^{-19}$\,J; gray area in the figure. However, a more accurate relation is obtained by observing that $A_{123}$ in Eq.~\ref{eq:Hamaker} goes roughly proportional to $(\varepsilon_3 - 1)/(\varepsilon_3 + 1)$  for systems with $\varepsilon_1 \gg \varepsilon_2$ and  $\varepsilon_3 \gtrsim \varepsilon_2$  $ = 1$. Assume now that the difference in the oxides' response behaviors is dominated by their 
respective value of the high-frequency dielectric constant. Then, the relation should be 
%
\begin{equation}
A_{123}(\varepsilon_3) \approx   
a_0\,\frac{(\varepsilon_\infty - 1)}{(\varepsilon_\infty + 1)}.
\label{eq:A123vseps}
\end{equation}
A regression analysis, employing $\varepsilon_\infty$ from Table~\ref{tab:eps}, results in $a_0 = 5.15 \times 10^{-19}$\,J for the Fe(liq)//oxide systems. This plain and manageable relation describes the trend of the Hamaker constant surprisingly well;  
see dashed black line in Figure~\ref{fig:results}(b). The relation equals correctly zero 
if the dielectric constant of the substrate would be one. 
Furthermore, we determine the value of $a_0$ with only the high-temperature phases of the oxides. 
However, since the relation should be applicable for any phase, we display also the data for SiO$_2$(tri) and ZrO$_2$(mono), represented by the smaller marks in the figure, demonstrating good behavior of Eq.~\ref{eq:A123vseps}.
\par
From Figure~\ref{fig:results}(b), one might think there should be a similar relation also for the Hamaker constant versus the static dielectric constant, $\varepsilon_0$, of the oxides. However, that is not applicable in our case, especially not for ZrO$_2$(tetra). The static dielectric constants are less important at high temperatures, and for two layers that are in close contact with each other.\cite{Sol2024} 
That is clear when comparing the results for ZrO$_2$(mono) and ZrO$_2$(tetra). The two phases have similar $A_{123}$ and comparable $\varepsilon_\infty$, but their $\varepsilon_0$ differ by a factor of about $2.5$.
In our case, $\varepsilon_0$ contributes only for the  first Matsubara frequency ($n = 0$), whereas the spectrum of the dielectric function relates to $\varepsilon_\infty$ in a much larger energy region.  
Hence, the main contribution to the Casimir-Lifshitz forces is, for the present systems and temperature, the electronic transitions [i.e, Eq.~\ref{eq:eps2c}(a)].
%
\par 
Wettability and stiction depend on the force between the liquid and the substrate, which is
calculated by the derivative of the work with respect to the distance, determined at the equilibrium distance $d_0$.
The resulting expression $F = A_{123}/6\pi d_0^3$ describes the force per area (alternatively, the pressure), 
where a positive value implies attraction.  
We do not find any straightforward relation between the force and the dielectric constant [Figure~\ref{fig:results}(c)].
Since our measured contact angles for calcia and zirconia  deviate from the literature data, we present 
with small marks in the figure also the results for the average values of their contact angles. Still, there is no clear trend in the relation. 
Whereas the Hamaker constant is independent of the auxiliary gap, which is evident by Eq.~\ref{eq:Hamaker}, 
the forces (as well as work of adhesion) depend inversely on $d_0$. The relation between $d_0$ and $\varepsilon_\infty$ is however less straightforward.
For instance, one can notice that zirconia has the smallest force (and smallest $W$; Table~\ref{table:HamakerFe}), 
despite having the largest dielectric constant which in turn implies the largest Hamaker constant according to Figure~\ref{fig:results}(b). 
The reason is that its auxiliary gap is quite large which weakens the cross-gap interactions.
Another example is alumina versus calcia. The have comparable high-frequency dielectric constants, but their forces deviate significantly.
Hence, one cannot draw conclusions about the wettability or stiction  based only on the size of the Hamaker constant.
Instead, one has to analyze also the contact angle in relation to the auxiliary gap. 
\par
In order to model the wettability of the liquid/ substrate systems, one should first describe the relation between the material properties and the contact angle $\theta$. We do not find any straight relation between the Hamaker constant and the contact angle (Figure~\ref{fig:results}(d), left axis), nor between the dielectric constant and the contact angle (right axis in the figure). That is true also when considering the average values of the measured contact angles $\theta_{ave}$ for CaO and ZrO$_2$; the smaller marks in the figure.
Hence, one has to incorporate the auxiliary gap $d_0$ in the description of the cross-interface interaction, i.e. by modeling $W$ versus $\theta$ together with a description of $d_0$ versus $\theta$.
\par 
One can in 
Table~3
observe that the work of adhesion is declining with increasing contact angle. That is expected, because the energy to separate the liquid droplet from the surface should be smaller for lower wettability. 
The dependence is almost linear with $W \approx 4.097 - 0.026\,\theta$\,J/m$^2$; gray area in Figure~\ref{fig:results}(e). 
However, the correct relation is $W = \gamma_1\cdot(\cos{\theta} + 1)$, which is displayed by the black dashed line in the figure. This relation goes to zero as the contact angle reaches $180^\circ$, i.e. for completely non-wetting  when the droplet is not attracted to the surface. Moreover, the maximum theoretical value of the work of adhesion is $2\gamma_1 = 3.68$\,J/m$^2$, which would occur at $0^\circ$ when the liquid completely wets the substrate and form a thin uniform layer. 
\par 
As discussed, the work of adhesion depends on the Hamaker constant, which is not an unambiguous material parameter unless accompanied by a consistent determination of the auxiliary gap. The gap is even more important when describing the cross-interface force as it is inversely proportional to $d_0^3$. It is therefore crucial to understand and correctly model $d_0$ when describing  interfacial properties as stiction of molten iron on the refractory nozzle wall, particle clogging of the nozzle, or oxide particle-particle agglomeration in the melt. Here, $d_0$ is described as an auxiliary distance because it is depending on how the interface is described by the three-layer model (Eq.~\ref{eq:Hamaker}), but also because $d_0$ is in the order of 0.1\,nm (1\,\AA). The diameters of atoms are typically $0.1$--$0.3$\,nm (e.g. the Ar atom has a diameter of $\sim$0.15\,nm),\cite{Slater1964} and $d_0$ with can therefore not be associated with a macroscopic layer of physical air or an argon gas. Certainly, sole atoms could penetrate into the interfacial layer, however, that would have neglectable impact on the Casimir-Lifshitz forces. 
Also, treating that effect would imply modeling of the atomic polarizability as well as including surface and interface relaxation of Fe(liq) and the oxides, and that is beyond the present macroscopic model of the material systems. Instead, it is an uttermost advantage with the present methodology to be able to accurately explore the cross-interface interactions with a macroscopic description of the layered systems, combined with an atomistic calculation of bulk material properties. 
\par
With the computed Hamaker constants 
(Table~3), 
it is clear that the auxiliary gap typically increases when the wettability becomes lower, i.e. for larger contact angles; see blue marks and right axis in Figure~\ref{fig:results}(e). That is reasonable, because a weaker interfacial interaction should yield a larger gap. 
Even though it is preferable with an explicit calculation of the Hamaker constant and its corresponding auxiliary gap, it would be useful if one could predict $d_0$ and $\theta$ for any new oxide compound or related alloys. 
From Eq.~\ref{eq:d0} it is clear that the gap depends on the square root of the Hamaker constant. 
Since $A_{123}$ does not deviate too much between the oxides, a very rough estimate would be to use an average value of it and use $d_0^2 \approx \overline{A}_{123}/12\pi\gamma_1(\cos{\theta} + 1)$. 
A simple linear regression yields $\overline{A}_{123} = 2.79 \times 10^{-19}$\,J; this is shown as the blue dotted line in Figure~\ref{fig:results}(e). Although that relation describes $d_0$ versus $\theta$ fairly well, 
it has the disadvantage of assuming that all liquid/oxide systems are associated with the same Hamaker constant. An improved approach would be to utilize the relation between the Hamaker constants and the high-frequency dielectric constants in Eq.~\ref{eq:A123vseps}, which then yields 
%
\begin{equation}
d_0(\theta,\gamma_1,\varepsilon_3)  \approx 
\sqrt{a_0\,\frac{(\varepsilon_\infty - 1)}{(\varepsilon_\infty + 1)} \frac{1} {12 \pi\, \gamma_1\, (\cos\theta + 1)}}.
\label{eq:d0vstheta} 
\end{equation}
This approximated expression generates actually very accurate auxiliary gaps as function of the contact angle; see red filled circles in Figure~\ref{fig:results}(e). It is applicable for our measured contact angles (blue marks) as well as for the average angles (the two smaller blue marks). The model involves however the constant $a_0$, which depends on the specific liquid, in our case molten iron. It would be of interest to know how much the value of $a_0$ deviates for different liquid/oxide systems.
%
%
\par
In addition to the work on the Fe(liq)/oxide systems, we present a brief analysis of the wettability of a liquid tin-bismuth alloy on the oxides.  This nontoxic alloy exhibits the advantage of becoming liquid at relatively low temperature, and can therefore be subject for experimental investigations of the fluid flow with direct measurement techniques.  
%
%
With this complementary study it is possible to judge the accuracy of our applied methodology for two rather different types of liquid metals.
\par
We use a similar approach to model the atomic configuration of liquid Sn-Bi as for liquid Fe, however now using a face-centered cubic Sn$_3$Bi structure that mimics an alloy with $\sim 37$\,wt\% Bi.
The wettability of the Sn$_3$Bi(liq)/oxides is modeled at $T = 473$\,K  (200\,$^\circ$C),\cite{LChen2024}
at which the liquid surface tension is $\gamma_1 = 0.46$\,J/m$^2$.\cite{Moser2001,Plevachuk2011}
For SiO$_2$ and ZrO$_2$, we consider the trigonal and monoclinic phases, respectively.
Due to the strong hygroscopic properties of CaO, the results for Sn$_3$Bi(liq)/CaO are not reliable for the present measurement conditions, and we therefore exclude that system.
For the four other refractory oxides the resulting Hamaker constants, contact angles, auxiliary gaps, and works of adhesion are presented in Table~\ref{table:HamakerSnBi}. 
One notices that the parameters are quite comparable to the corresponding values for Fe(liq)/oxides, however, $A_{123}$ are smaller and $d_0$ are typically larger. 
As a consequence, $W$ is smaller. This is expected, as the surface tension of Sn$_3$Bi(liq) is about four times smaller than that of Fe(liq). Therefore, the wettability in terms of the contact angle should be relatively high, and that is also what the sessile drop measurements manifest.
\\
%
\begin{table}[!h]
\caption{ Similar to Table~\ref{table:HamakerFe}, 
but here for the Sn$_3$Bi(liq)-on-oxide and at $T = 473$\,K.}
\renewcommand{\arraystretch}{1.2}
\begin{tabular}{lcccc}
   \hline  \hline 
 Oxide & $A_{123}$ [$10^{-19}$\,J] &  $\theta$ [$^\circ$] &$d_0$\,[nm] & $W$ [J/m$^2$]\\
   \hline
  Al$_{2}$O$_{3}$ & 2.38 &           110\,$\pm 2$   & 0.144 & 0.31 \\
  MgO             & 2.24 & \phantom{0}86\,$\pm 3$   & 0.109 & 0.50 \\
  SiO$_{2}$(tri)  & 1.87 & \phantom{0}80\,$\pm 1$   & 0.096 & 0.54 \\
  ZrO$_{2}$(mono) & 2.86 &           101\,$\pm 3$   & 0.142 & 0.38 \\
  \hline \hline
\end{tabular}
\label{table:HamakerSnBi}
\end{table}
\par
\begin{figure}[!ht]
 \centering
  \includegraphics[width=0.88\columnwidth]{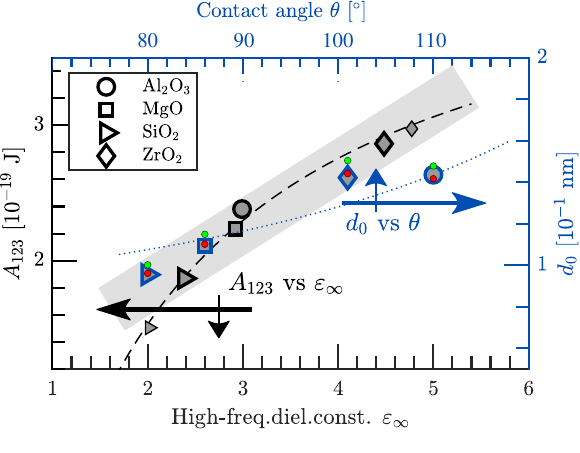}
  \caption{Corresponding graphs as in Figures~\ref{fig:results}(b) and (e), but for the Sn$_3$Bi(liq)//oxides and at $T = 473$\,K. 
  Black small marks are the high-temperature phases SiO$_2$(cub) and ZrO$_2$(tetra).  
  Red and green filled circles represent Eq.~\ref{eq:d0vstheta} with $a_0 = 4.59$ and $5.15\times 10^{-19}$\,J, respectively.}
\label{fig:resultsSnBi}
\end{figure}

Likewise the Fe(liq)/oxide systems, the Hamaker constants as function of the high-frequency dielectric constant are parameterized according to Eq.~\ref{eq:A123vseps}. The result for the Sn$_3$Bi(liq)/oxides is $a_0 = 4.59\times 10^{-19}$\,J, which generates the black dashed line in Figure~\ref{fig:resultsSnBi}. The corresponding behavior of $d_0$ versus $\theta$, described by Eq.~\ref{eq:d0vstheta} (red filled circles in the figure), is as good as for the Fe(liq)/oxides, whereas using an average value of the Hamaker constant fails to a large extent (blue dotted line).
Actually, the relation in Eq.~\ref{eq:d0vstheta} (based on Eq.~\ref{eq:A123vseps}) holds with a reasonable accuracy for all considered liquid/oxide systems with one and the same constant. To exemplify that statement, the green filled circles in Figure~\ref{fig:resultsSnBi} are the results when using $a_0 = 5.15\times 10^{-19}$\,J that was obtained for the Fe(liq)/oxide systems.
Further, a regression analysis that includes data from both liquids  yields $a_0 = 5.0\times 10^{-19}$\,J. This fit could serve as a crude approximation to be valid for the two molten metals considered here, all five refractory oxides, and for angles ranging at least $80^\circ \lesssim \theta \lesssim 140^\circ$. The temperature dependence (see Eq.~\ref{eq:Hamaker}) has only a moderate impact, as it is compensated by the step lengths of the Matsubara frequencies. 
However, we want to accentuate that an accurate and more secure determination of the cross-gap interactions requires the proper and consistent modeling of the Casimir-Lifshitz free energies.
%
%
\section{Conclusions} 
\label{sect:summary}
In this study, we are exploring the wettability of liquid iron on the five most common refractory oxides, namely Al$_{2}$O$_{3}$, CaO, MgO, SiO$_{2}$, and ZrO$_{2}$, utilizing a sessile drop technique and first-principles modeling of the cross-interface interactions. The crystal structures of these oxides are analyzed through room-temperature X-ray, verifying the relaxed structures from the DFT calculations. Thereafter, the complex dielectric functions of these compounds are computed, and the dielectric constants for both the room-temperature phases and the high-temperature phases are presented. Moreover, we model the dielectric response function of liquid Fe at $T = 1823$\,K, whereupon confirming that the spectrum agrees with available experimental data. This model of Fe(liq) significantly simplifies the theoretical exploration of the liquid's behavior under high-temperature conditions. By combining these atomistic DFT calculations with the macroscopic Casimir-Lifshitz forces, we calculate the cross-interface interactions for the Fe(liq)/oxide systems at $1823$\,K, introducing the auxiliary gap $d_0$ of the interlayer to link the work of adhesion $W$ to the Hamaker constant $A_{123}$. The macroscopic three-layer system, denoted Fe(liq)//oxide, is foundational for describing also particle-particle and particle-nozzle interactions in order to simulate agglomeration and nozzle clogging in steel casting processes. It is generic in the sense that it can be applied to other macroscopic liquid/solid system. Complementing the macroscopic view, the DFT calculations come into play, aiming at a microscopic description of these interactions. 
These calculations provide thereby an in-depth perspective of the intermolecular interactions between Fe(liq) and the refractory oxides. The resulting estimate of the Hamaker constants for the material systems is a pivotal step in comprehending their interfacial interactions. We explain that $A_{123}$ depends predominantly on the high-frequency dielectric constant, $\varepsilon_\infty$, although a full description of the dielectric functions is required for an accurate calculation. Employing only the static dielectric constants in a simplified model of the Hamaker constant would result in incorrect conclusions. 
With the Casimir-Lifshitz forces, and specifically the Hamaker constants, the wettability of the Fe(liq)/oxide systems is explored by means of the Young's model. That equation bridges the contact angle, an indicative measure of the wettability, with the interfacial energies of the solid, liquid, and vapor phases. The contact angle data is measured with a sessile drop method at 1823\,K, which provides an empirical complement and support to the theoretical models. 
\par
\begin{figure}[!ht]
 \centering
  \includegraphics[width=0.85\columnwidth]{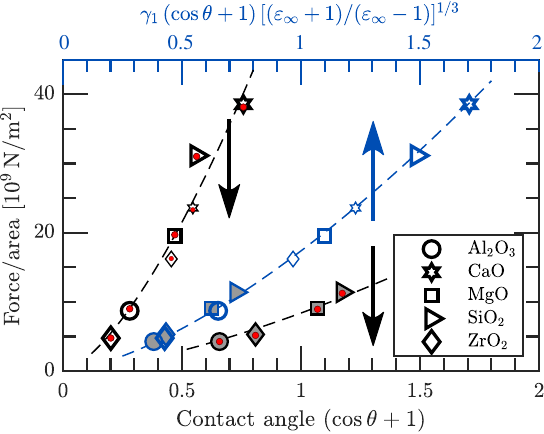}
  \caption{
  Forces per unit area for the Fe(liq)//oxides (open marks) and the Sn$_3$Bi(liq)//oxides (gray filled marks). Black dashed lines indicate that the forces go roughly as $(\cos\theta+1)^{3/2}$.
  Red circles include Eq.~\ref{eq:d0vstheta} with $a_0 = 5.15$ and $4.59\times 10^{-19}$\,J for Fe(liq) and Sn$_3$Bi(liq), respectively. Blue dashed line is the best fit to Eq.~\ref{eq:A123vseps} for both liquids, i.e. $a_0 = 5.0\times 10^{-19}$\,J and by incorporating also the liquids' surface tension in the model: $\gamma_1 = 1.843$ and $0.460$\,J/m$^2$, respectively. 
  }
\label{fig:resultsFall}
\end{figure}

A key finding of our research is the variance in the value of the spatial auxiliary gap, $d_0$, for the five systems at $T = 1823$\,K, indicating differences in the cross-interface interactions. We emphasize that a consistent model of the auxiliary gaps is needed to accompany the Hamaker constants.
The reason is that the relevant physical property for the considered material system is actually the work of adhesion, which involves both the Hamaker constant and the gap, i.e. $W \propto A_{123}/d_0^2$. Utilizing the Hamaker constants in models of particle-on-nozzle stiction or particle-particle agglomeration requires therefore also a proper determination of $d_0$. By comparing the results for the Fe(liq)/oxide systems with the corresponding results for the Sn$_3$Bi(liq)/oxides, we demonstrate that this auxiliary gap depends not only on the oxide, but also on the type of liquid. Hence, a crude model of $d_0$, for instance constant value independent of the oxide, could yield inaccurate conclusions. However, for the two liquids and the five oxides there is a common characteristic. In Figure\,\ref{fig:resultsFall}, the forces are depicted for the Fe(liq)/oxides as well as for the Sn$_3$Bi(liq)/oxides. From $F = 2W/d_0 = 2\gamma_1(\cos\theta +1)/d_0$ and eq.~\ref{eq:d0}, the forces should behave approximately as $(\cos\theta +1)^{3/2}$. That is also what is obtained; see the two black dashed lines for Fe(liq) and Sn$_3$Bi(liq)  in the figure. Furthermore, if one incorporates eq.~\ref{eq:d0vstheta} and the liquids' surface tension in the model, the cross-interface forces can approximately be described as
%
\begin{equation}
F(\theta,\gamma_1,\varepsilon_3)  \approx 
\sqrt{48\pi\,\frac{1}{a_0}\frac{(\varepsilon_\infty + 1)}{(\varepsilon_\infty - 1)} \gamma_1^3\, (\cos\theta + 1)^3}\,.
\label{eq:Fvstheta} 
\end{equation}
This characteristic of this expression with $a_0 = 5.0\times 10^{-19}$\,J is represented by the blue dashed line in Figure~\ref{fig:resultsFall}.
%
\par
One reason that Eq.~\ref{eq:Fvstheta} with a constant $a_0$ is valid for all considered liquid//oxide systems in this study, is that Eq.~\ref{eq:A123vseps} can be applied for both liquids through $\varepsilon_1 \gg \varepsilon_2$ and  $\varepsilon_3 \gtrsim \varepsilon_2$. That is a reasonably good approximation for the dielectric functions up to $\sim 10$\,eV; see Figure~\ref{fig:dft_diel}(b). Thereby, the value of the Hamaker constant does not depend on other properties of the liquid metals, apart from the very large $\varepsilon_1$ due to the Drude contribution for metals. 
However, to describe chemical physics of the interfacial system, also the auxiliary gap has to be included.  
For example, the cross-interface forces in Eq.~\ref{eq:Fvstheta} depend on $d_0$, and that can be accounted for by the surface tension of the liquid, which in turn depends implicitly on the liquid's dielectric response. Thus, the simplified model of Eq.~\ref{eq:A123vseps}, omitting the liquid's property, is not enough to analyze wettability, agglomeration, etc. 
Another reason Eq.~\ref{eq:A123vseps} is valid with reasonable accuracy for all systems is that the oxides have comparable band-gap energies. Thereby, the gradually decline of the dielectric responses for energies above $\sim 6 $\,eV are similar for all oxides. To exemplify this relevance, we construct a hypothetical dielectric function which is similar to that of a true oxide, but associated with a smaller band-gap energy. 
More concretely, we calculate the dielectric function of Al$_2$O$_3$, employing the PBEsol functional but excluding the band-gap correction. The gap energy is then underestimated by 2.7\,eV. Smaller gap yields larger $\varepsilon_\infty$, and therefore the dielectric response is scaled by a constant so that $\varepsilon_\infty$ agrees with the value for true alumina. (For simplicity in this example, we neglect to correct the sum rule.) As a result, the calculated Hamaker constant of the hypothetical oxide is $2.53 \times 10^{-19}$\,J, which is $8$\% smaller than the value for true alumina. The error is perhaps not devastating for theoretically analyzing the wettability considering the accuracy of corresponding measurements, but it illustrates that one should handle the approximated relation in Eq.~\ref{eq:A123vseps} with care. 
\par
The insights gained in this work are particularly relevant for processes involving liquid Fe and the refractory materials where the wettability plays a significant role, specifically for optimizing steel casting and ceramic manufacturing. The synergy between the theoretical modeling and the experimental validation, as is exemplified in this study, underscores a robust approach for future studies in the field. Merging the DFT calculations into the modeling of the Casimir-Lifshitz dispersion forces is demonstrated to provide an accurate as well as accessible methodology in determining the interaction parameters, offering also an understanding of the complex relationships between molecular and macroscopic properties.  We advocate for further and extended research into the temperature-dependent properties of other material systems, which would enrich the understanding of material behaviors in extreme conditions. 
%
\section{Supporting Information} 
Supporting Information: 
Experimental setup of the furnace used for the contact angle measurements, measured X-ray diffraction patterns, and crystal structures considered for the DFT calculations. 
%
\begin{acknowledgments}
SK, BG, and CP acknowledge financial support from the European Union's Horizon 2020 research and innovation programmes for the project INEVITABLE with grant agreement No.~869815 and the project HIYIELD with grant agreement No.~101058694. 
They acknowledge access to high-performance computing resources via the National Academic Infrastructure for Supercomputing in Sweden (NAISS), provided by the National Supercomputer Centre (NSC) at Link\"{o}ping University and by the PDC Center for High Performance Computing at KTH Royal Institute of Technology. 
WC acknowledges support by UniversalLab Gmbh in Switzerland, and he thanks Mr. Tim Wang at the same laboratory for expertise help of the XRD data analysis and interpretation. MB's research contribution is part of the project No. 2022/47/P/ST3/01236 co-funded by the National Science Centre and the European Union's Horizon 2020 research and innovation programme under the Marie Sk{\l}odowska-Curie grant agreement No. 945339. MB's part of the work took place at "ENSEMBLE3 - Centre of Excellence for nanophotonics, advanced materials and novel crystal growth-based technologies" project (grant agreement No. MAB/2020/14) carried out within the International Research Agendas programme of the Foundation for Polish Science co-financed by the European Union under the European Regional Development Fund, the European Union's Horizon 2020 research and innovation programme Teaming for Excellence (grant agreement. No. 857543) for support of this work. MB's research contribution to this publication are created as part of the project of the Minister of Science and Higher Education 'Support for the activities of Centers of Excellence established in Poland under the Horizon 2020 program' under contract No. MEiN/2023/DIR/3797. 
\end{acknowledgments}
\bibliography{Reference}
\end{document}


\subsection{Experimental work}
\subsubsection{Sessile drop experiments}

    \begin{figure}
        \centering
        \includegraphics[width=1.0\linewidth]{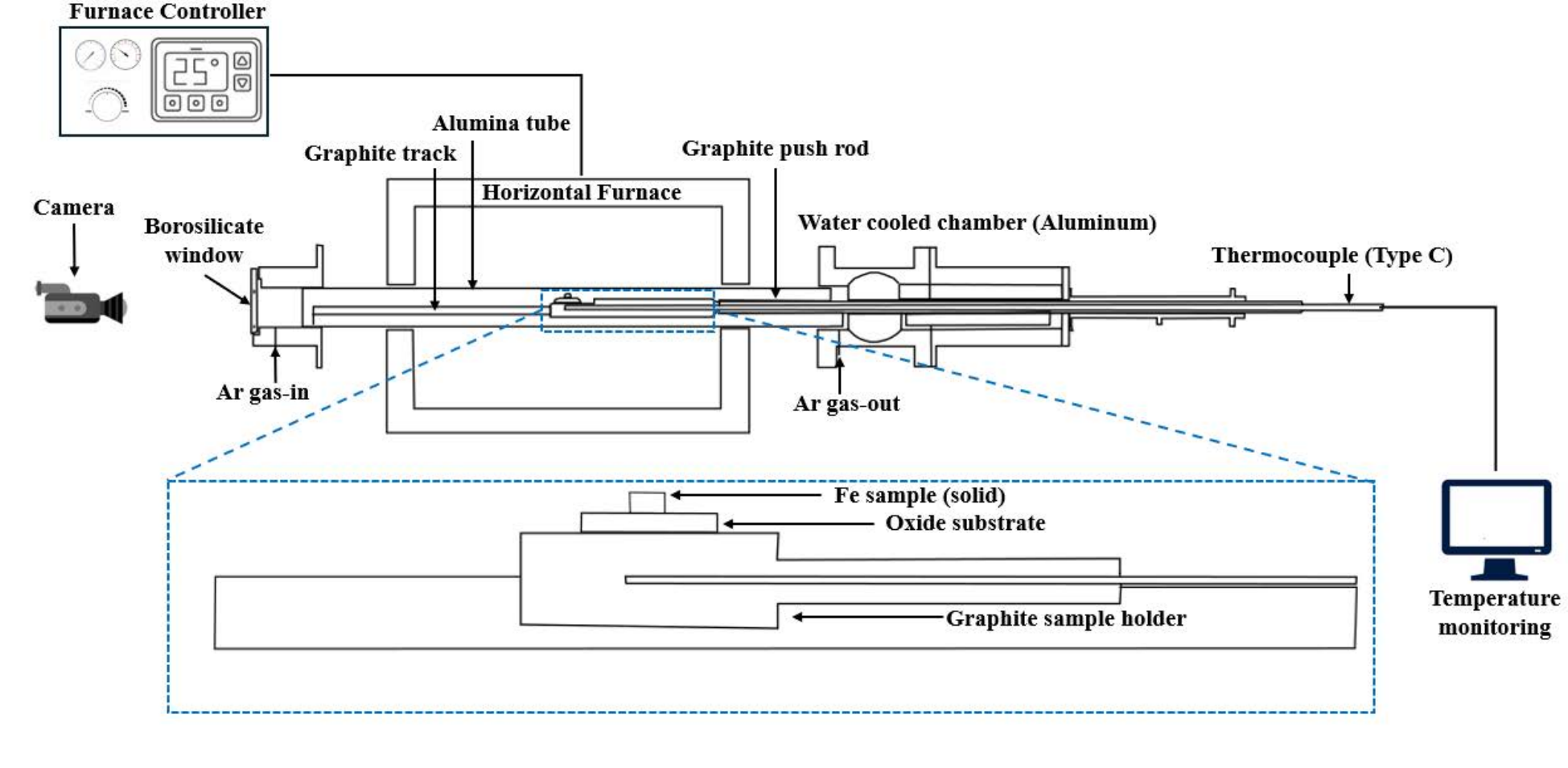}
        \caption{Experimental setup of the furnace used for contact angle measurement}
        \label{fig:experimental_setup}
    \end{figure}
    
    \begin{figure}
        \centering
        \includegraphics[width=1.0\linewidth]{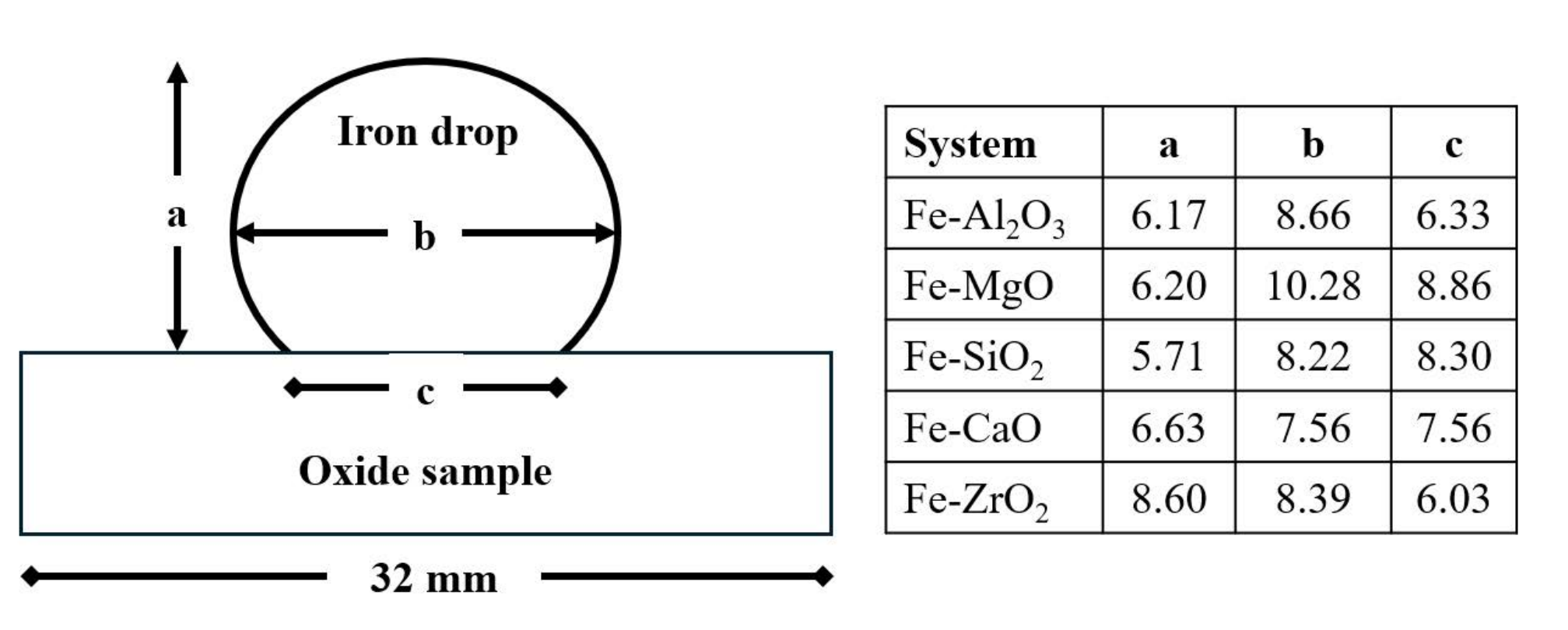}
        \caption{Three parameters describing the size of the Fe(liq) drop: {\bf a} represents the height, and {\bf b} is the maximum diameter of the drop, and {\bf c}  measures the contact length where the drop meets the oxide substrate; all parameters specified in millimeters. }
        \label{fig:drop_size}
    \end{figure}

\subsubsection{X-ray Diffraction analysis}
X-ray diffraction (XRD) analysis was specifically conducted on refractory oxide powder particles. The primary purpose of XRD in this study was to validate the lattice constants obtained from the DFT calculations. The XRD machine used for the analysis was bruker d2 phaser.

\begin{figure}
    \centering
    \includegraphics[width=0.5\linewidth]{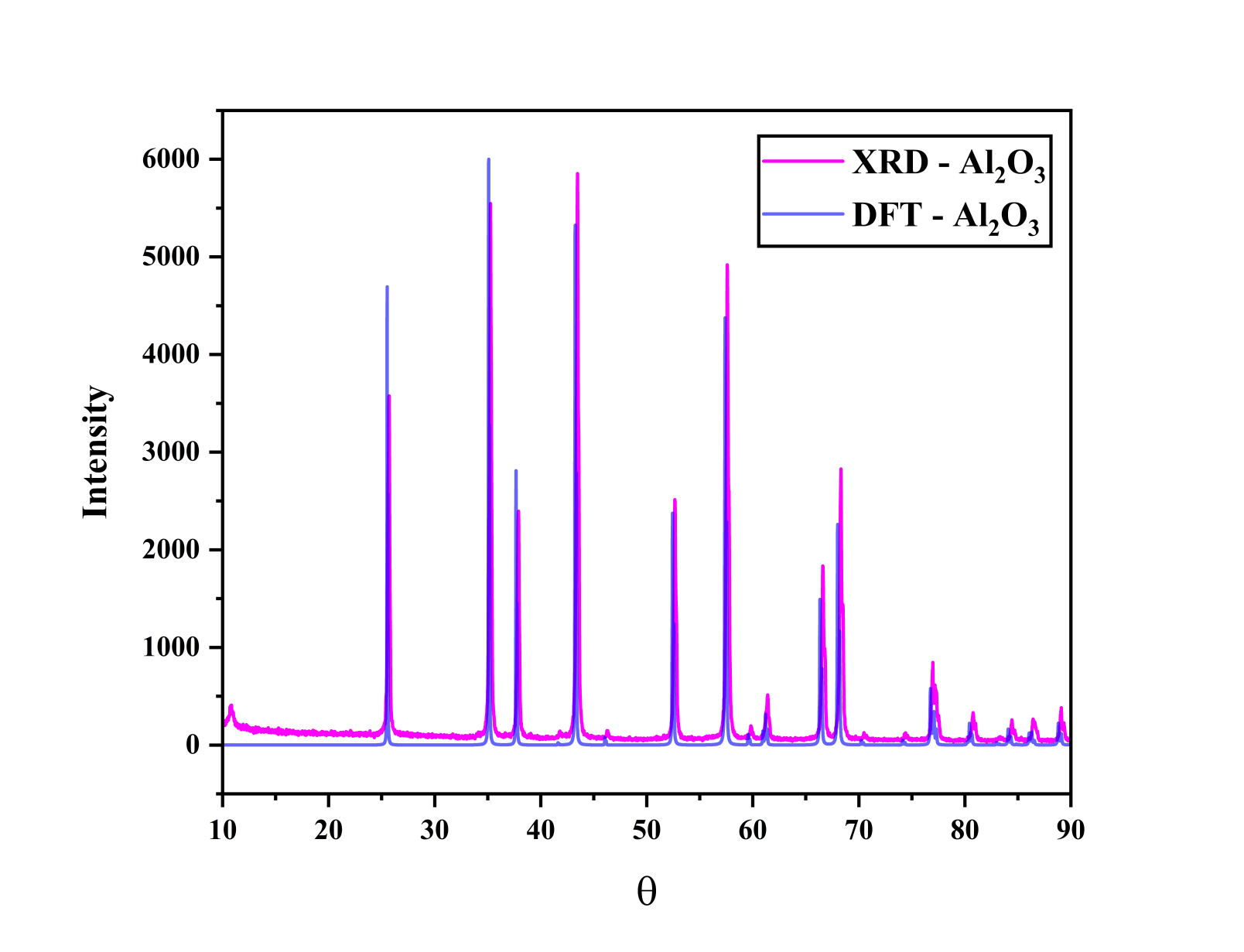}
    \caption{Powder diffraction pattern for Al$_{2}$O$_{3}$ from XRD measurements and DFT structure}
    \label{fig:XRD_dft_al2o3}
\end{figure}

\begin{figure}
    \centering
    \includegraphics[width=0.5\linewidth]{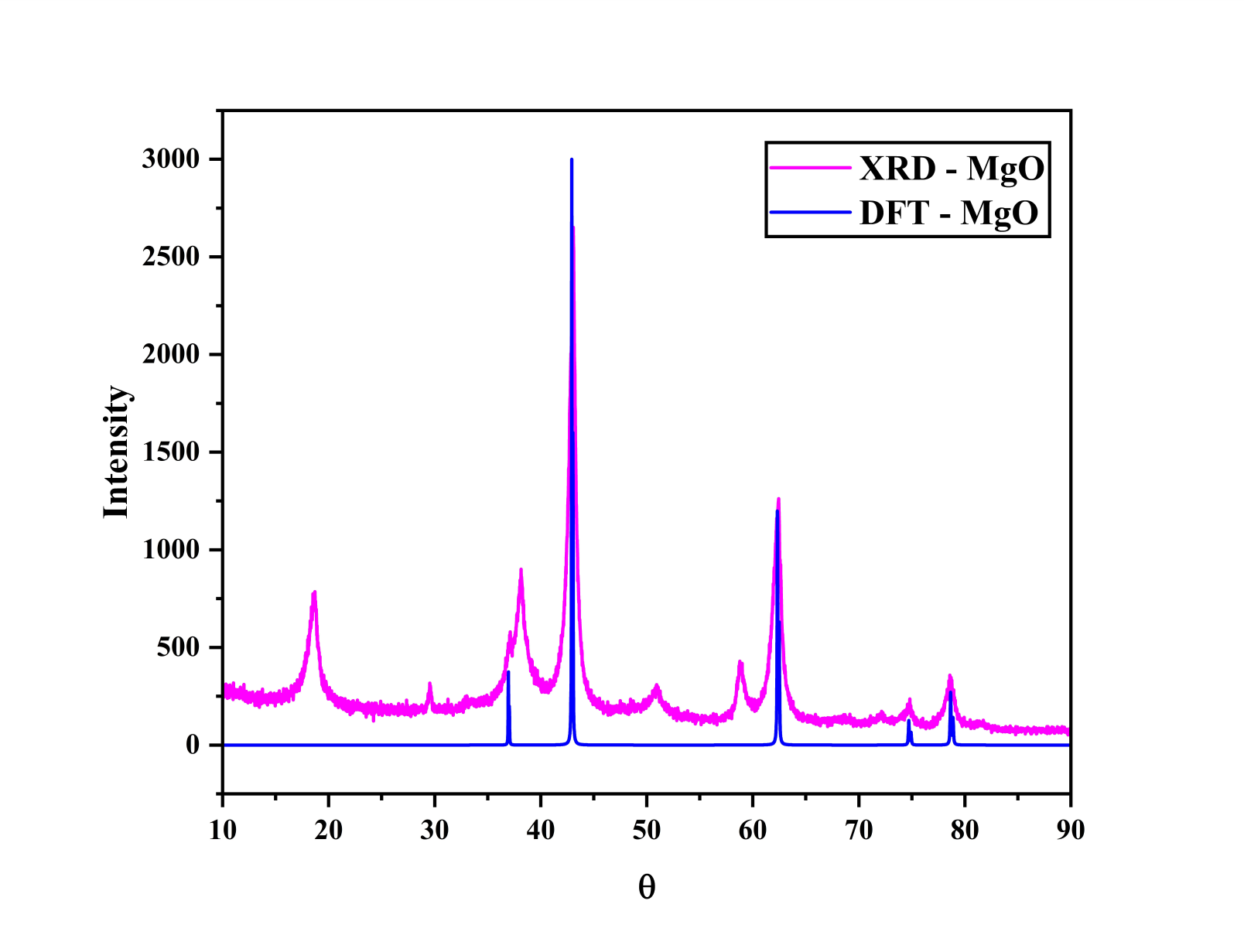}
    \caption{Powder diffraction pattern for MgO from XRD measurements and DFT structure}
    \label{fig:XRD_dft_MgO}
\end{figure}

\begin{figure}
    \centering
    \includegraphics[width=0.5\linewidth]{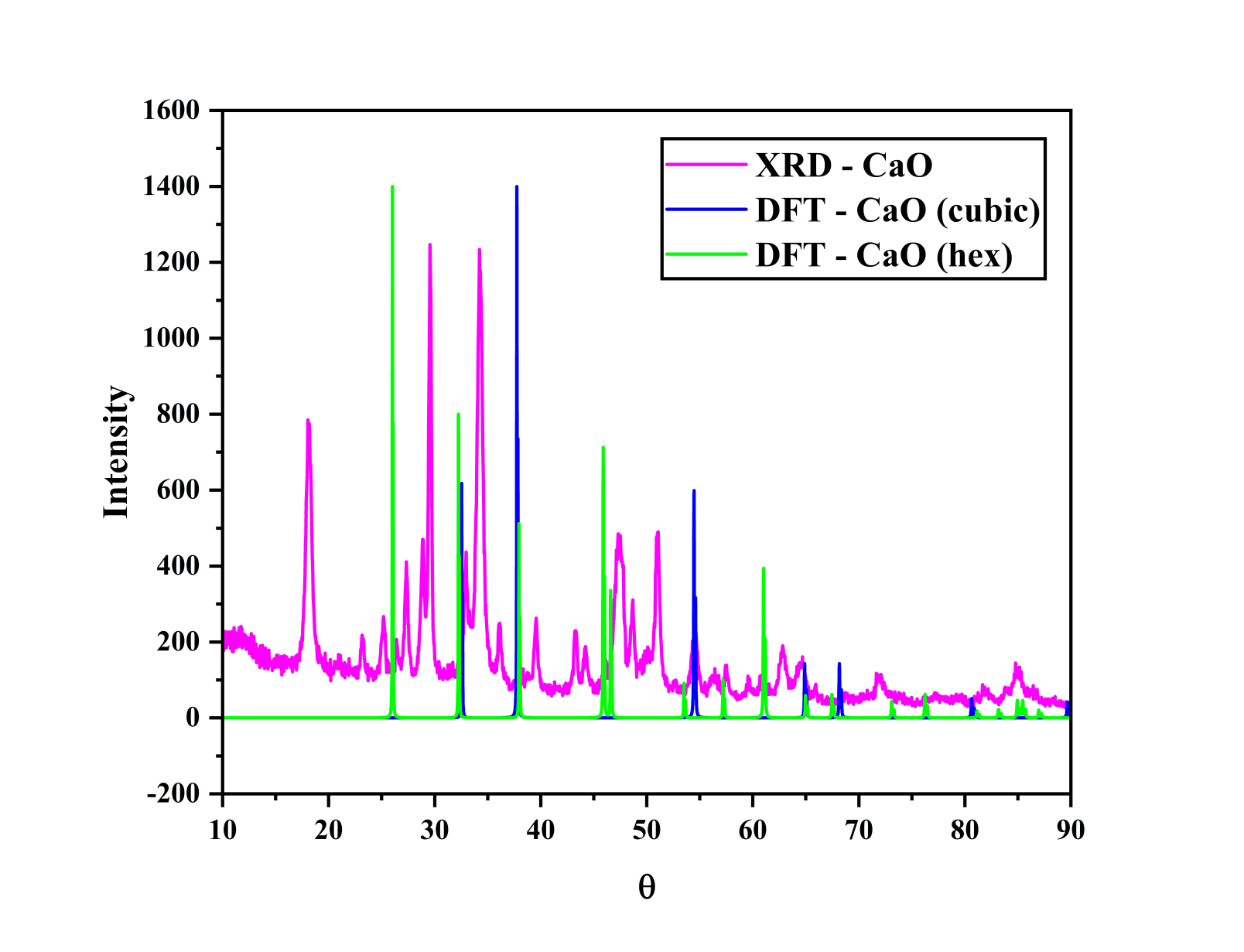}
    \caption{Powder diffraction pattern for CaO from XRD measurements and DFT structure}
    \label{fig:XRD_dft_CaO}
\end{figure}

\begin{figure}
    \centering
    \includegraphics[width=0.5\linewidth]{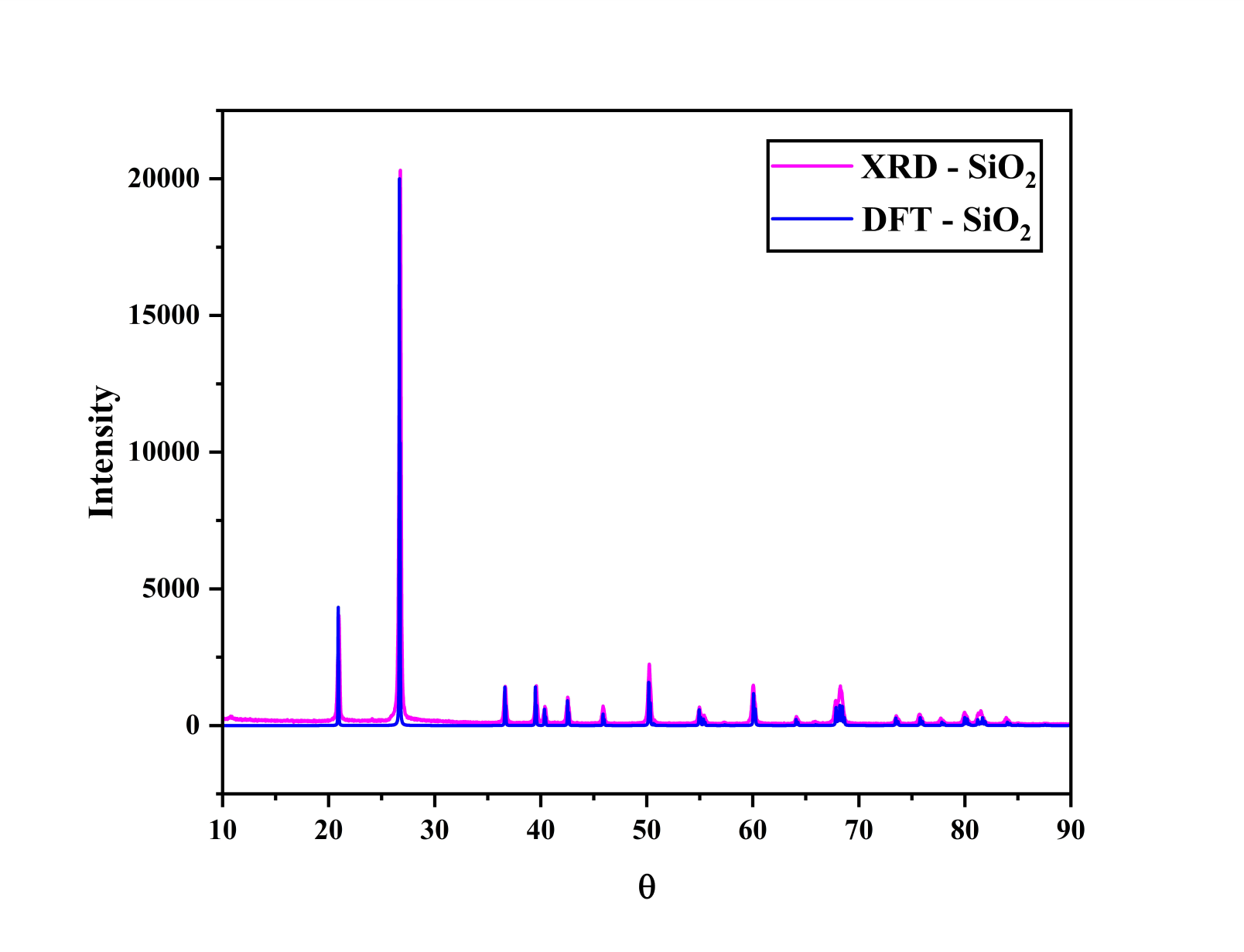}
    \caption{Powder diffraction pattern for SiO$_{2}$ from XRD measurements and DFT structure}
    \label{fig:XRD_dft_SiO2}
\end{figure}

\begin{figure}
    \centering
    \includegraphics[width=0.5\linewidth]{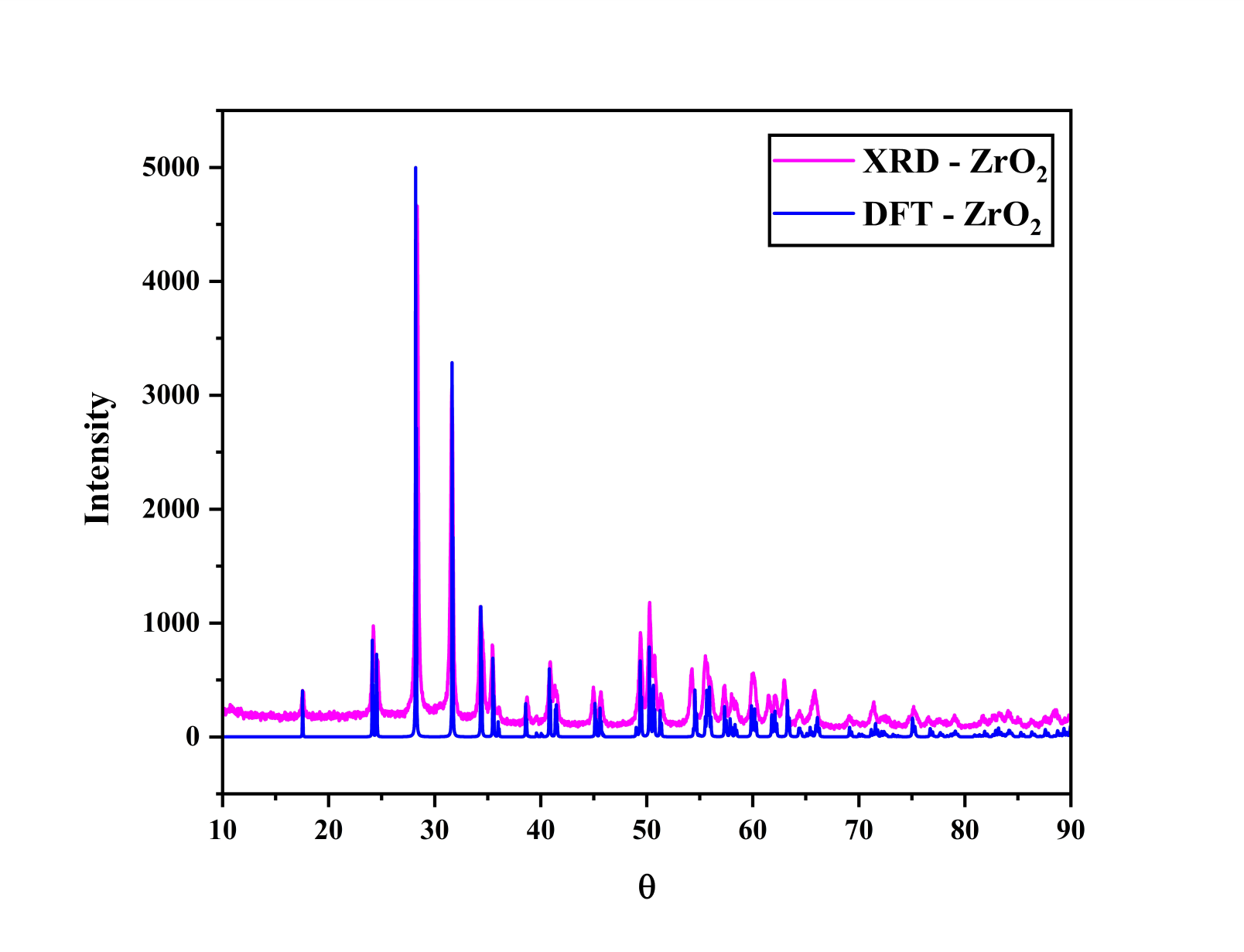}
    \caption{Powder diffraction pattern for ZrO$_{2}$ from XRD measurements and DFT structure}
    \label{fig:XRD_dft_ZrO2}
\end{figure}

\subsubsection{Structures considered for the DFT calculations}
Figures have been generated with the software package Visualization for Electronic and STtructural Analysis; K. Momma and F. Izumi, "VESTA 3 for three-dimensional visualization of crystal, volumetric and morphology data," J. Appl. Crystallogr., 44, 1272--1276 (2011).
\begin{figure}
    \centering
    \includegraphics[width=0.5\linewidth]{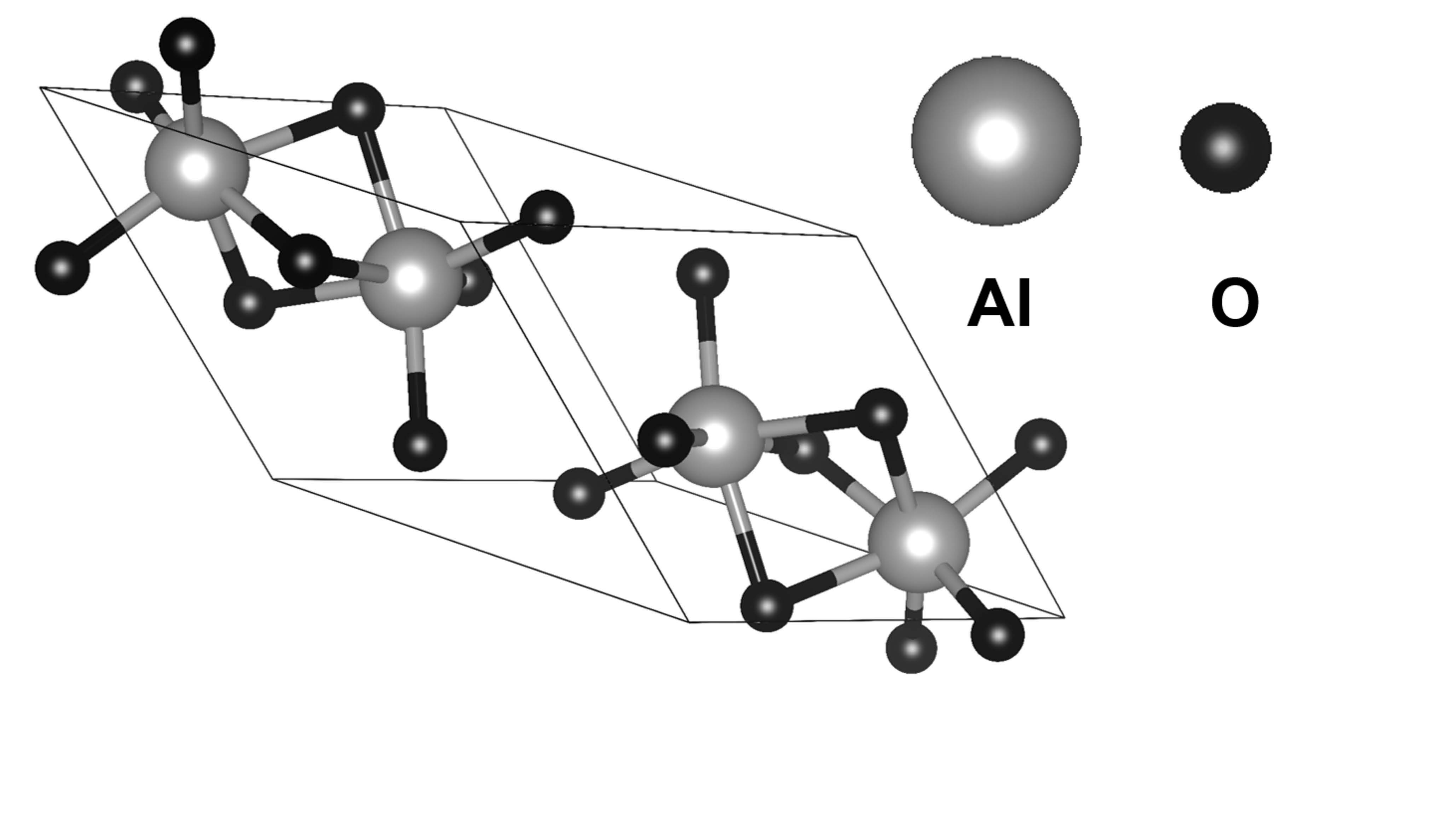}
    \caption{Structure of Al$_{2}$O$_{3}$ considered for the DFT calculations}
    \label{fig:dft_al2o3}
\end{figure}

\begin{figure}
    \centering
    \includegraphics[width=0.5\linewidth]{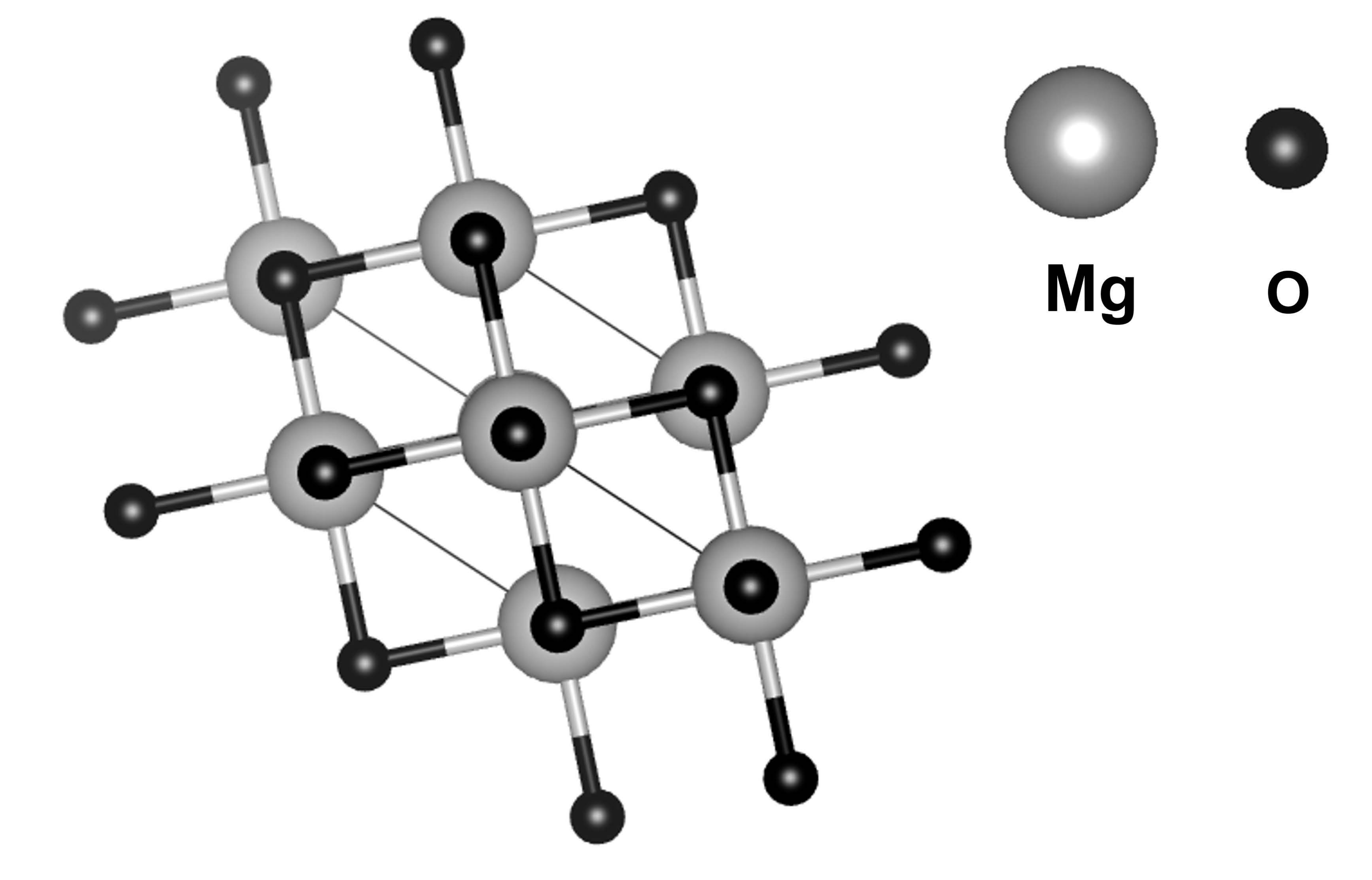}
    \caption{Structure of MgO considered for the DFT calculations}
    \label{fig:dft_mgo}
\end{figure}

\begin{figure}
    \centering
    \includegraphics[width=0.5\linewidth]{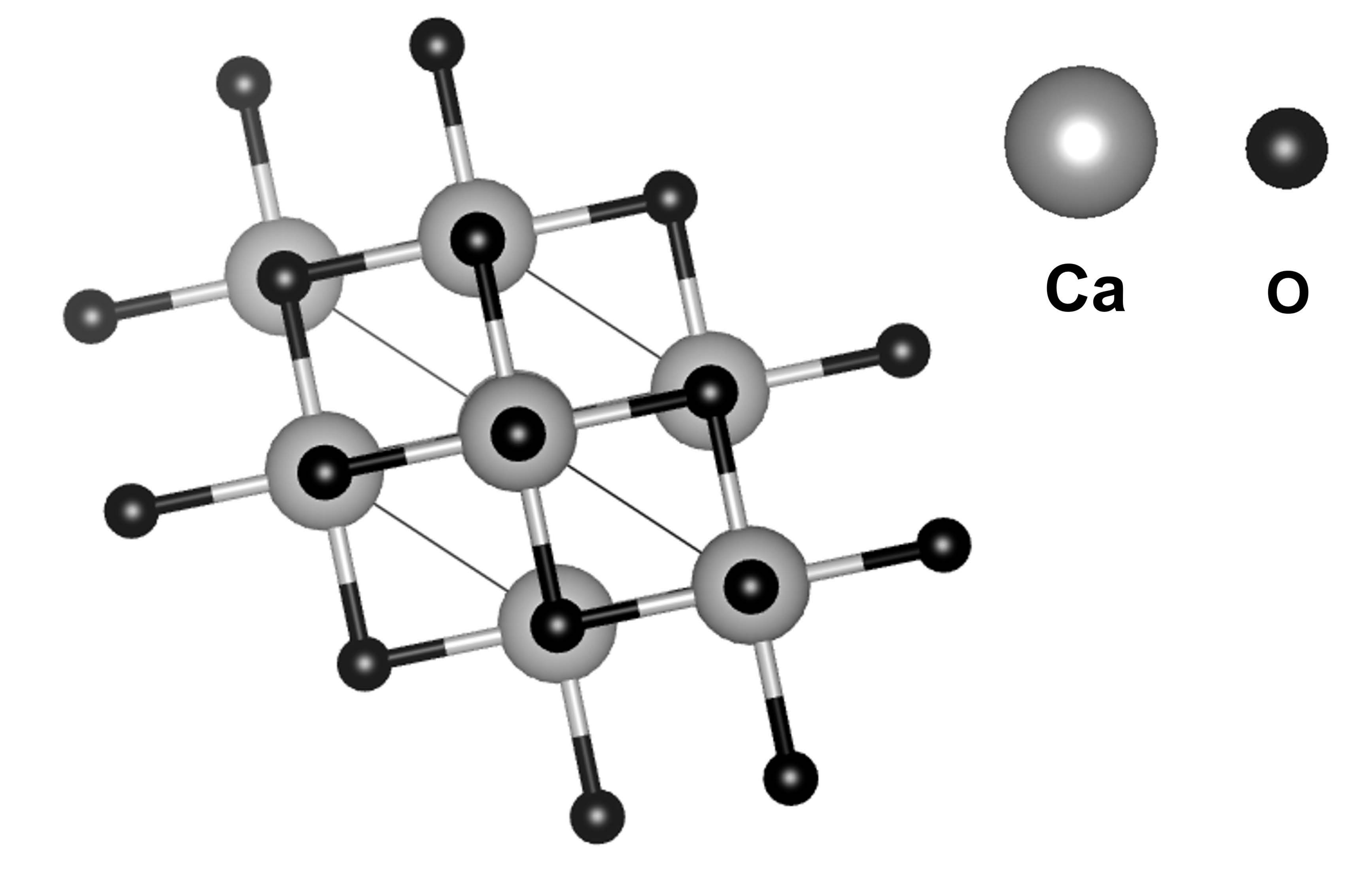}
    \caption{Structure of CaO considered for the DFT calculations}
    \label{fig:dft_cao}
\end{figure}

\begin{figure}
    \centering
    \includegraphics[width=0.5\linewidth]{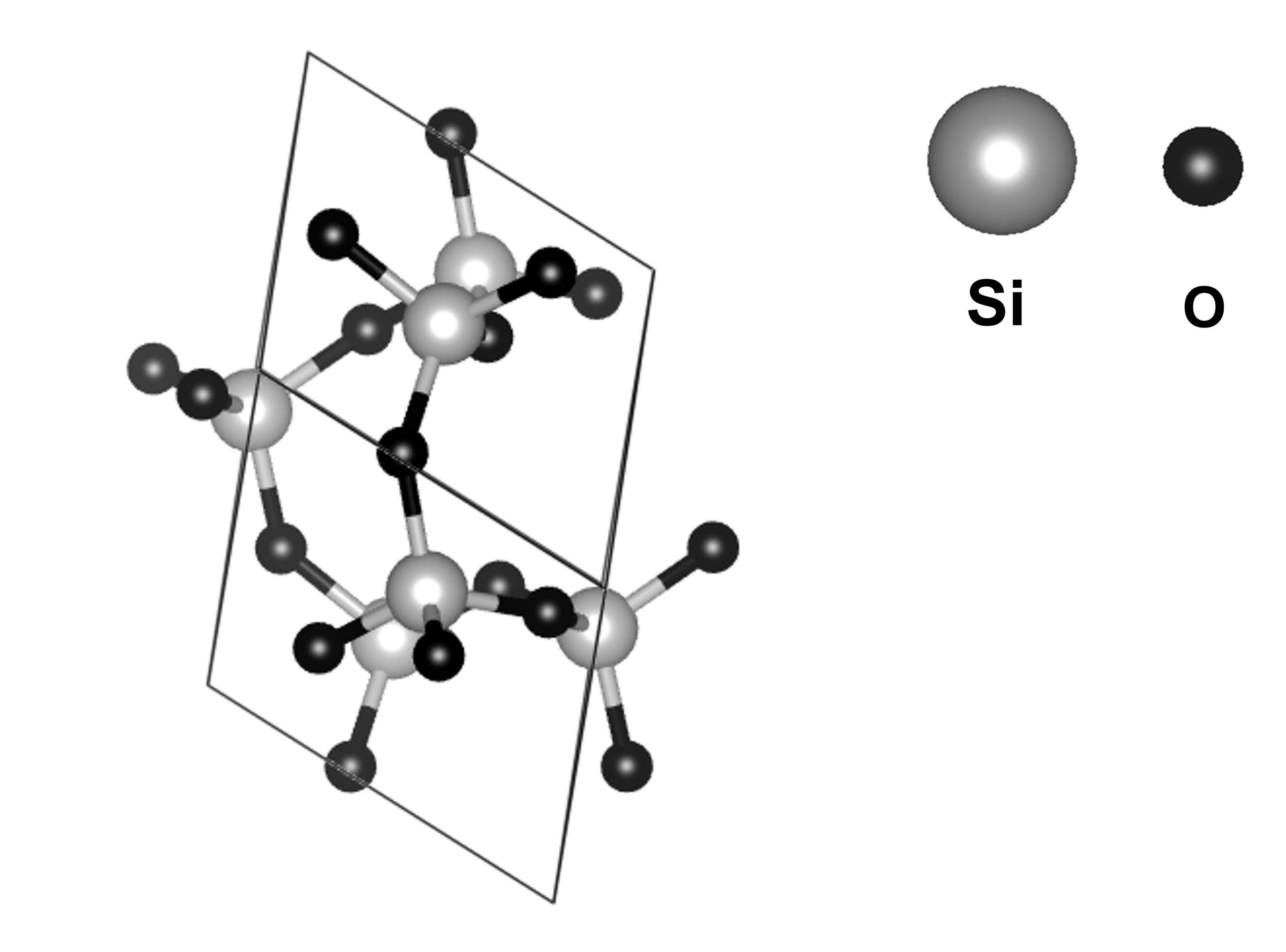}
    \caption{Structure of SiO$_2$ considered for the DFT calculations}
    \label{fig:dft_sio2}
\end{figure}

\begin{figure}
    \centering
    \includegraphics[width=0.5\linewidth]{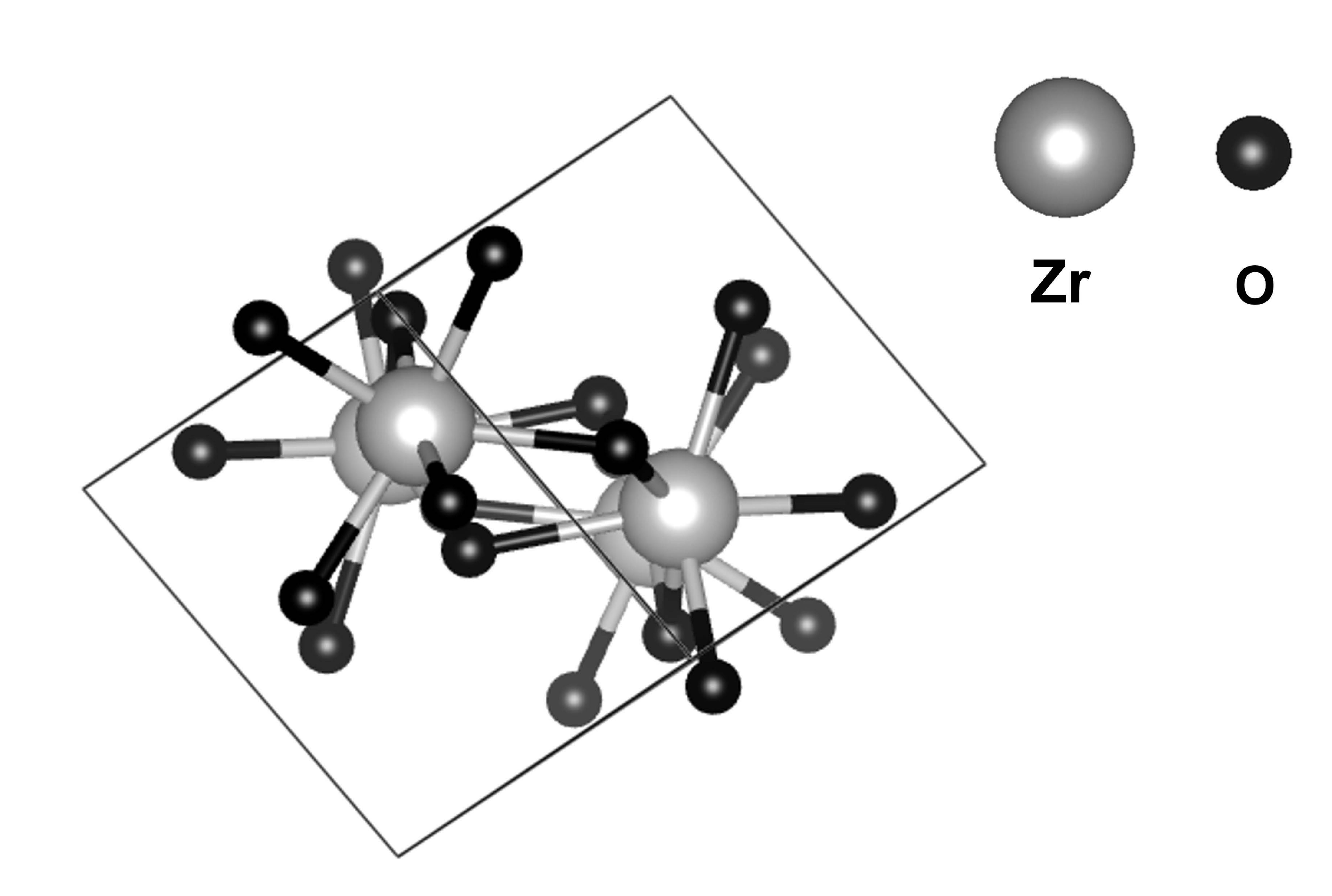}
    \caption{Structure of ZrO$_2$ considered for the DFT calculations}
    \label{fig:dft_zro2}
\end{figure}